\documentclass [12pt]{article}
\usepackage {graphicx}
\usepackage{epsfig,amsmath,latexsym,amssymb}
\usepackage{bm}

\usepackage{booktabs}
\usepackage{multirow}
\begin{document}


\title{Markov chain Monte Carlo estimation of default and recovery: dependent via the latent systematic factor}

\author{Xiaolin Luo$^{1}$ \quad Pavel V.~Shevchenko$^{2,\ast}$}

\date{\footnotesize{Working paper, 1st version 5 November 2010; this version 10 April 2013}}

\maketitle

\begin{center}
\footnotesize { \textit{$^{1}$ CSIRO Mathematics, Informatics and Statistics, Sydney, Australia; e-mail: Xiaolin.Luo@csiro.au \\
$^{2}$ CSIRO Mathematics, Informatics and Statistics, Sydney,
Australia;
e-mail: Pavel.Shevchenko@csiro.au  \\
$^*$ Corresponding author} }
\end{center}

\begin{abstract}
\noindent  It is a well known fact that recovery rates tend to go
down when the number of defaults goes up in economic downturns. We
demonstrate how the loss given default model with the default and
recovery dependent via the latent systematic risk factor can be
estimated using Bayesian inference methodology and Markov chain
Monte Carlo method. This approach is very convenient for joint
estimation of all model parameters and latent systematic factors.
Moreover, all relevant uncertainties are easily quantified.
Typically available data are annual averages of defaults and
recoveries and thus the datasets are small and parameter uncertainty
is significant. In this case Bayesian approach is superior to the
maximum likelihood method that relies on a large sample limit
Gaussian approximation for the parameter uncertainty. As an example,
we consider a homogeneous portfolio with one latent factor. However,
the approach can be easily extended to deal with non-homogenous
portfolios and several latent factors.

\vspace{1cm} \noindent \textbf{Keywords:} parameter uncertainty,
probability of default, loss given default,  economic capital,
Markov chain Monte Carlo, Bayesian inference, credit risk
\end{abstract}

\pagebreak

\section{Introduction}
\label{sec:introductionords} 
\begin{figure}[!htbp]
\centerline{\includegraphics[scale=0.9]{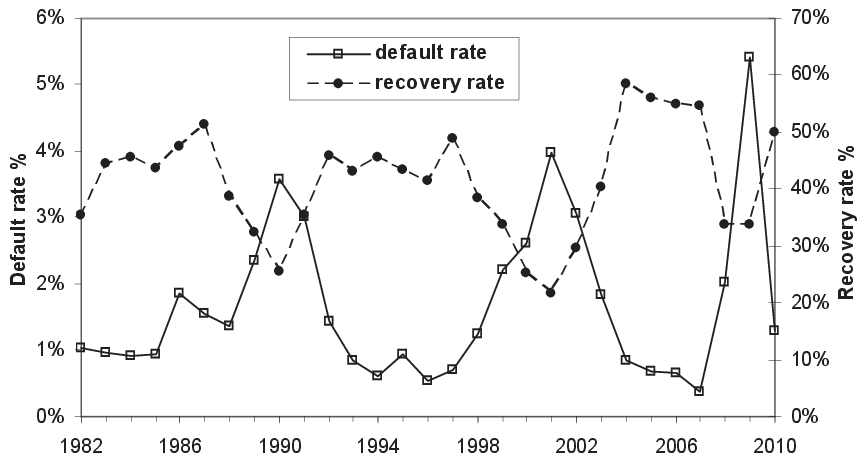}} \caption{Global
corporate default and recovery annual rates from Moody's (2011)}
\label{data_fig}
\end{figure}

Default and recovery rates are key
components of Loss Given Default (LGD) credit risk models. The
classic LGD model implicitly assumes that the default rates and
recovery rates are independent \nocite{bluhm02} (Bluhm \textit{et
al} 2002). There is empirical evidence that recovery rates tend to
go down just when the number of defaults goes up in economic
downturns that is clearly observed in historical data in Figure
\ref{data_fig}. Motivated by this fact, Frye
(2000a)\nocite{Frye_April2000}\nocite{Frye00},
 Pykhtin (2003)\nocite{Pykhtin} and
D\"{u}llmann and Trapp (2004)\nocite{DuellmannTr04} extended the
classic model to include systematic risk in recovery rates,
incorporating a non-zero correlation between default rates and
recovery rates  driven by the systematic factor. They considered
three extensions to
 account for the systematic risk in recovery
rates under three different assumptions for the distribution of
recovery rates: Frye (2000a)\nocite{Frye_April2000}\nocite{Frye00}
-- a normal distribution; Pykhtin (2003) -- a log-normal
distribution; D\"{u}llmann and Trapp (2004)\nocite{DuellmannTr04}
and \nocite{Scho01}
 Sch\"{o}nbucher (2001) -- a
logit-normal distribution.  The extended models are still
parsimonious, yet they represent an important enhancement of credit
risk models used in earlier practice, for example,
\nocite{CreditMetr} CreditMetrics (Gupton \textit{et al} 1997)  and
\nocite{CreditRisk} CreditRisk+ (Credit Suisse Financial Products
1997) that do not account for systematic risk factor driving both
default and recovery rates. Other extensions considering the
correlation between  risk drivers of default and recovery are found
in \nocite{Cantor05} Cantor and Varma (2005) and \nocite{Scheule05}
R\"{o}sch and Scheule (2005). These models (among others) have been
suggested by some banks for assessment of the Basel II ``downturn
LGD" requirement, see Basel Committee on Banking Supervision
(2005)\nocite{BaselPar468_2005}. The Basel II ``downturn LGD"
reasoning is that recovery rates may be lower during economic
downturns when default rates are high; and that a capital should be
sufficient to cover losses during these adverse circumstances. For a
good review of credit risk LGD models, see Altman
(2006)\nocite{altman2006default}.

D\"{u}llmann and Trapp (2004)\nocite{DuellmannTr04} summarized the
empirical literature on systematic risk in recovery rates, and found
a broad agreement that default rates and business cycle are
correlated. They calculated the maximum likelihood estimators (MLEs)
of model parameters for the default and recovery rate distributions;
estimated the correlations of default and recovery rates with the
systematic risk factor; and found that economic capital (EC),
defined as the 0.999 quantile of the annual loss distribution, is
significantly higher in the extended LGD models (in comparison with
the classic one-factor model) due to dependence of recoveries on the
systematic risk factor. It was also observed that EC estimates are
very close to each other for all three distributional assumptions
for the recovery rates.

Publicly available data provided by Moody's or Standard\&Poor's
rating agencies  are annual averages of defaults and recoveries.
These data are of limited size, covering a couple of decades at
most. For example, in the study of D\"{u}llmann and Trapp
(2004)\nocite{DuellmannTr04}, the default and recovery data have
eighteen points covering an eighteen-year period 1982-1999.
Inevitably the limited data size could pose significant instability
and uncertainty in  the LGD model parameter estimates. None of the
various studies, including the extension work of Frye (2000a,
2000b)\nocite{Frye_April2000}\nocite{Frye00}, Pykhtin
(2003)\nocite{Pykhtin} and
 D\"{u}llmann and Trapp (2004)\nocite{DuellmannTr04} specifically addressed the
quantitative impact of parameter uncertainty. Increasingly,
quantification of parameter uncertainty has become a key component
of financial risk modeling and  management. Recent examples of
addressing parameter uncertainty in operational risk and insurance
include \nocite{LuShDo07} Luo \textit{et al} (2007) and
\nocite{PeShWu09} Peters \textit{et al} (2009a).

Bayesian inference is a convenient approach to jointly estimate all
model parameters and latent factors, and all relevant uncertainties.
It is especially useful when data are limited and parameter
uncertainty is large. In this case Bayesian approach is superior to
the maximum likelihood method that relies on a large sample limit
Gaussian approximation for the parameter uncertainty. Under the
Bayesian approach, the inference is based on the distribution of the
parameters and latent factors given data (so-called posterior
distribution). Typically, the posterior distribution is not
available in closed-form but can be easily estimated numerically
using Markov chain Monte Carlo (MCMC) method. In this paper, we
demonstrate how the extended LGD model can be estimated using
Bayesian inference and MCMC method. For illustration, we consider
homogeneous portfolio with one latent factor. However, the approach
can be easily extended to non-homogeneous portfolios and several
latent factors.

The organization of this paper is as follows. Section 2 first
describes the credit risk model setup, particularly the extended
default and recovery models considered by  Frye
(2000a)\nocite{Frye_April2000}, Pykhtin (2003)\nocite{Pykhtin} and
 D\"{u}llmann and Trapp (2004)\nocite{DuellmannTr04}. This is followed by
a discussion on various EC estimates and the corresponding
algorithms, both for the finite number of borrowers and for the
limiting case of the infinitely granular portfolio. The emphasis is
on how to account for parameter uncertainty using Bayesian inference
and MCMC. Section 3 presents the likelihood functions for the LGD
model. This includes the full joint likelihood for default and
recovery as well as two-stage approximation used in Frye
(2000b)\nocite{Frye00} and D\"{u}llmann and Trapp
(2004)\nocite{DuellmannTr04}. For the latter, we derive the
closed-form MLEs for the recovery process parameters in addition to
the known closed-form MLEs for the default parameters. Section 4
describes the Bayesian inference formulation and the MCMC simulation
algorithm for the posterior distribution of the LGD model
parameters. Sections 5 and 6 present MCMC results in comparison with
the MLEs using annual default and recovery rates for corporate
bonds. Results in Section 5 are for the 1982-1999 data period, the
same time period as studied in D\"{u}llmann and Trapp
(2004)\nocite{DuellmannTr04}, while results in Section 6 are for the
period 1982-2010 covering the recent global financial crisis.
Concluding remarks are given in the final section.


\section{LGD Model}

The standard one-factor LGD model assumes a homogenous loan
portfolio where the distribution of its loss vector that collects
losses of individual loans is exchangeable (invariant) under
permutations of its components. Following Frye
(2000a)\nocite{Frye_April2000}, Pykhtin (2003)\nocite{Pykhtin} and
 D\"{u}llmann and Trapp (2004)\nocite{DuellmannTr04}, the key
characteristics of the one-factor model are summarized as follows.

Consider a portfolio of $J$ borrowers (firms) over a chosen time
horizon. To avoid cumbersome notation, we assume that the $j$th
borrower has one loan with principal amount $A_j$. The loss rate
(loss amount relative to the loan amount) of the portfolio due to
defaults  is

\begin{equation}\label{portfolioLoss_eq}
L=\sum_{j=1}^J w_jL_j=\sum_{j=1}^J w_jI_j \max(1-R_j,0),
\end{equation}

\noindent where we have the following definitions.
\begin{itemize}
\item $w_j$ is the weight of loan $j$ in the portfolio,
$w_j=A_j/\sum_{m=1}^JA_m$.
\item $L_j$ is the loss rate of loan $j$ due to potential default.
\item $1-\max(1-R_j,0)=\min(R_j,1)$ is the  recovery rate of loan $j$ after
default.
\item $I_j$ is an indicator variable associated with the default of
firm $j$, $I_j=1$ if firm $j$ defaults, otherwise $I_j=0$.
\end{itemize}

Quantity $R_j$ can be loosely interpreted as the value of collateral
per unit of exposure (e.g. see Frye 2000a\nocite{Frye_April2000}).
When this quantity exceeds 1 (i.e. the value of collateral exceeds
the value of exposure), then $100\%$ recovery is assumed. In general
$R_j$ is not the same as recovery rate since the latter is subject
to a cap of 1.

Following D\"{u}llmann and Trapp (2004)\nocite{DuellmannTr04}, in
this study we do not explicitly impose the restriction
  $0\leq R_j \leq 1$. In fact, results in D\"{u}llmann and Trapp (2004)\nocite{DuellmannTr04} show that the unbounded normal distribution for recovery rate gives a capital estimate very close  to that given by the
  properly bounded logit-normal distribution (the relative difference is less than $1\%$).
  As in D\"{u}llmann and Trapp (2004)\nocite{DuellmannTr04}, for simplicity
  the term ``recovery rate" is  used for the quantity
  $R_j$ in the rest of the paper.

\vspace{0.5cm} \noindent {\it \bf Remark:} The above notation is for
a given time period. Later, starting from Section 3, we consider the
model over a number of time periods $t=1,2,\ldots ,T,T+1$ that will
add index $t$ to all random variables. Here, $T+1$ refers to the
next year. It is assumed that all random variables involved in the
model are independent between different time periods. However, the
model can be easily extended to have explicit time dependence.

\subsection{Modeling Default}

  Denote the probability of default for firm $j$ by $p$, i.e. $\mathrm{Pr}[I_j=1]=p$. Let  $C_j$ be an underlying latent  random
variable  such that firm $j$ defaults if $C_j<\Phi^{-1}(p)$, where
$\Phi(\cdot)$  is the standard normal distribution and
$\Phi^{-1}(\cdot)$ is its inverse. That is, $I_j=1$ if
$C_j<\Phi^{-1}(p)$ and $I_j=0$ otherwise. $C_j$ describes the
overall financial condition (financial well-being) of firm $j$ over
a time horizon. The value $C_j$ for each firm depends on a
systematic risk factor $X$ and a firm specific (idiosyncratic) risk
factor $Z^C_j$ as

\begin{equation}
C_j=\sqrt{\rho}X+\sqrt{1-\rho}Z^C_j,
\end{equation}

\noindent where $Z^C_1,\ldots ,Z^C_J$ are all independent. Also, $X$
and $Z^C_j$ are assumed to be independent and from the standard
normal distribution.

Conditional on $X$, the financial conditions of any two firms are
independent. The parameter $\rho$ quantifies the extent of exposure
of a firm's asset value to the fluctuations in the business cycle.
Unconditionally,  it  measures the correlation between financial
conditions of two firms. The value of $\rho\in [0,1]$ is assumed to
be the same for all firms but can be extended to be firm specific if
required.

\subsection{Modeling Recovery}
The extended LGD models proposed and studied by D\"{u}llmann and
Trapp (2004)\nocite{DuellmannTr04}, Frye (2000a,
2000b)\nocite{Frye_April2000}\nocite{Frye00} and Pykhtin
(2003)\nocite{Pykhtin}
  account for
systematic risk in  recovery rates under three different assumptions
for the distribution of recovery rates. Define

\begin{equation}
 V_j=\mu +\sigma\sqrt{\omega}X+\sigma\sqrt{1-\omega}Z_j,
\hspace{0.3cm} \omega\in[0,1],
\end{equation}

\noindent where $X$ and $Z_j$ are assumed to be independent and from
the standard normal distribution, and parameter $\omega$ is
restricted to the interval $[0,1]$. Also, $Z_j$ and $Z_j^C$ are
assumed independent too. Note, the one-factor model in Pykhtin
(2003) allows for correlation between $Z_j$ and $Z_j^C$. The three
models for the recovery rate are then defined through $V_j$ as
follows.

\begin{itemize}
\item The first extended model, as initially suggested by Frye (2000a)\nocite{Frye_April2000},
assumes a normal distribution for the recovery rates, i.e. the
recovery rate $R_j$ of loan $j$ is given by
\begin{equation} \label{eqn_recovery}
R_j=V_j.
\end{equation}

An advantage of the above model is that parameters $\mu$ and
$\omega$ directly represent the mean and correlation of recoveries
respectively.

\item The second extension, initially proposed by \nocite{Scho01}
 Sch\"{o}nbucher (2001), assumes that the recovery rate $R_j$ follows a
logit-normal distribution, i.e.

\begin{equation}
R_j=\frac{\exp(V_j)}{1+\exp(V_j)}.
\end{equation}

The above model satisfies the restriction $0 < R_j < 1$.

\item The third model, following
Pykhtin (2003)\nocite{Pykhtin}, has a log-normal distribution for
the recovery rate
\begin{equation}
R_j=\exp(V_j).
\end{equation}

\end{itemize}

The study by D\"{u}llmann and Trapp (2004)\nocite{DuellmannTr04}
shows that EC estimates from the above three recovery models are
very close to each other; only about $2\%$ difference exists among
the EC values estimated by these models. In addition, they carried
out Shapiro-Wilk test and Jarque-Bera test for normality, and found
that the normal distribution assumption for the recovery rate is
favored by the p-values over the other two models. Thus in the
present study, we will concentrate on the first recovery model given
by (\ref{eqn_recovery}), i.e. we assume a normal distribution for
the
 recover rate, but it is not difficult to use other recovery distributions.
 Another reason for our choice of model
(\ref{eqn_recovery}) is because we do not have the original data for
individual recoveries but only the average recovery rates; and we
can use the fact that the distribution of the average of normally
distributed independent random variables is still normal.

\subsection{Economic Capital }
Following the literature, we define the  economic capital (EC) as
the 0.999 quantile of the distribution of loss $L$ defined in
(\ref{portfolioLoss_eq}). Specifically, the quantile $Q_q$ is
defined as

\begin{equation}
 Q_q({\bm\theta })\equiv Q_q =  \inf \{z:\Pr [L > z|\bm\theta] \le 1 - q\}= \inf \{z:F_L(z|\bm\theta)\geq
 q\},
\label{eqn_quantile}
\end{equation}

\noindent where $q$ is a  quantile level (e.g. 0.999);
$F_L(z|\bm\theta)$ is distribution function of the random loss $L$;
the corresponding density of $L$ is denoted as $f_L(z|\bm\theta)$;
and ${\bm\theta }=(p,\rho,\mu,\sigma,\omega)$ are the model
parameters.

There are  different ways of estimating this high quantile, some are
based on point estimates (e.g. MLEs) of parameters and others
account for parameter uncertainty.
 We are interested in comparing these different estimates
of the quantile and quantifying the impact of different assumptions,
particularly the impact of parameter uncertainty.

\subsubsection{Quantile point estimates}
For a given model with parameters ${{\bm \theta }}$, the quantile
$Q_q({\bm\theta })$ is a function of
 ${{\bm \theta }}$. Typically, given observations, the MLEs ${{\widehat{\bm\theta }}}$
are used as the ``best fit'' point estimates for ${{\bm \theta }}$.
Then, the  loss density for the next time period is estimated as
$f_L(z\vert {{\widehat {\bm\theta }}})$ and its  quantile is
estimated as $Q_q({{\widehat {\bm\theta }}})$. In general, the
distribution of $L$ is not
 tractable in closed form for an arbitrary portfolio. In this case, Monte Carlo method for
 simulating $L$ in (\ref{portfolioLoss_eq})  for given parameters ${\bm\theta }$
 can be used as follows.

 \vspace{0.3cm} \noindent {\bf  Algorithm 1} ({\bf Quantile given
 parameters})

\begin{enumerate}
  \item Draw a single independent sample from the standard normal distribution for the
systematic factor $X$.
  \item For each borrower $(j=1,\ldots ,J)$, draw an independent sample from the standard normal distribution for
the idiosyncratic default risk factor $Z^C_j$;  calculate $C_j$ as
in (2); and let $I_j=1$ if $C_j<\Phi^{-1}(p)$ and $I_j=0$ otherwise.
  \item Draw an independent sample from the standard normal distribution for
the idiosyncratic recovery factor $Z_j$ and calculate $R_j=\mu
+\sigma\sqrt{\omega}X+\sigma\sqrt{1-\omega}Z_j$.
  \item Find loss $L$ for the entire portfolio using (\ref{portfolioLoss_eq}). This
is  a sample from the loss distribution $F_L(\cdot|\bm\theta)$.
\item Repeat steps 1-4 to obtain $N$ samples
of $L$ with $N$ sufficiently large for high quantile calculations
(i.e. numerical error due to finite number of simulations is small
enough).
\item Estimate $Q_q({\bm \theta})$ using obtained samples of $L$ in
the standard way (e.g. using sample with the index $\lceil Nq\rceil$
after sorting in the ascending order).
\end{enumerate}

In practice, the parameters ${{\bm \theta }}$ are unknown and it is
important to account for this uncertainty when the quantile is
estimated, especially in the case of small datasets. A standard
frequentist approach to estimate this uncertainty is based on
limiting results of normally distributed MLEs for large datasets.
Then information matrix (calculated from the second order
derivatives of the likelihood) is used to estimate the covariances
between MLEs. In this paper we take Bayesian approach, because
dataset is small and the distribution of parameter uncertainty is
very different form normal. Estimation of the quantile accounting
for parameter uncertainty under the Bayesian inference framework
will be discussed in Section 4.

\subsubsection{Economic capital under the limiting condition}
In the case of a diversified portfolio with a large number of
borrowers, the idiosyncratic risk can be eliminated and the loss
depends on $X$ only. \nocite{Gordy02} Gordy (2002) has shown that
the distribution of portfolio loss $L$ has a limiting form as
$J\rightarrow\infty$, provided that each weight $w_j$ goes to zero
faster than $1/\sqrt{J}$. The limiting loss rate $L^\infty$ is given
by the expected loss rate conditional on the systematic factor $X$

\begin{equation} \label{eqn_infL0}
L^\infty\equiv L^\infty(X)=E[L|X]=\sum_{j=1}^J
w_jE[L_j|X]=\sum_{j=1}^J w_jE[I_j\max(1-R_j,0)|X],
\end{equation}
\noindent i.e. the limiting loss $L^\infty$ is just a function of
$X$ and the distribution of $L^\infty$ is fully implied by the
distribution of $X$.

Conditional on  $X$, the default indicator variable $I_j$ and the
recovery rate $R_j$ are independent because $Z^C_j$ in (2) and $Z_j$
 in (3) are independent. Thus, the limiting loss (\ref{eqn_infL0}) for $J\rightarrow\infty$ becomes

\begin{equation} \label{eqn_infL}
L^\infty=\sum_{j=1}^J w_jE[I_j|X] E[\max(1-R_j,0)|X]=\sum_{j=1}^J
w_j\Lambda_j(X)  S_j(X),
\end{equation}

\noindent where $\Lambda_j(X)=E[I_j|X]$ is the conditional
  probability of default of firm $j$ and
$S_j(X)=E[\max(1-R_j,0)|X]$ is the conditional expected value of
loss rate, both are functions of $X$.

 Bank loans are subject to the borrower specific risk
and systematic risk. The former can be controlled or even
neutralized by diversification. Note that
(\ref{eqn_infL0}-\ref{eqn_infL}) is valid for a non-homogeneous
portfolio. For a homogeneous portfolio, probability of default and
recovery rates (or loss given default) are not firm specific, i.e.
$\Lambda_j(X)=\Lambda(X)$ and $S_j(X)=S(X)$ for all $j$, and
(\ref{eqn_infL}) simplifies to

\begin{equation} \label{eqn_infL2}
L^\infty=\sum_{j=1}^J w_j\Lambda(X) S(X)=\Lambda(X)
S(X)=L^\infty(X).
\end{equation}

\noindent That is, the limiting loss rate of the diversified
homogenous portfolio is a function of $X$ only.   As in the model
underlying the internal ratings-based risk weights of Basel II, EC
is determined with the assumption that the bank loan portfolio is
fully diversified and EC is only held for systematic credit risk.
Because $L^\infty(X)$ is a monotonic decreasing function of random
variable $X$, and $X$ is from the standard normal distribution, the
quantile of $L^\infty(X)$ at level $q$, can be calculated as
$$Q^\infty_q =L^\infty\left(X=\Phi^{-1}(1-q)\right).$$

 As  in D\"{u}llmann
and Trapp (2004)\nocite{DuellmannTr04}, we define EC  of the
diversified portfolio loss distribution $L^\infty(X)$ as the $0.999$
quantile
\begin{eqnarray} \label{eqn_ec}
EC^{\infty} &= &Q^\infty_{0.999} =L^\infty\left(\Phi^{-1}(0.001)\right) \nonumber \\
&=&\Lambda\left(\Phi^{-1}(0.001)\right)\times
S\left(\Phi^{-1}(0.001)\right) =\mathrm{PD}\times \mathrm{LGD},
\end{eqnarray}

\noindent where
$$\mathrm{PD}= \Lambda(\Phi^{-1}(0.001))\quad
\mbox{and}\quad \mathrm{LGD}=S(\Phi^{-1}(0.001) $$ are stressed
probability of default (\emph{stressed PD})
 and stressed loss given default (\emph{stressed LGD}) respectively.
The  stressed PD can be inferred from the observed default rates; it
is determined once the unconditional probability of default $p$ and
parameter $\rho$ are estimated. Using (2), the conditional
probability of default can be written as a function of $X$

\begin{equation}
\Lambda(X)=\Phi
\left(\frac{\Phi^{-1}(p)-\sqrt{\rho}X}{\sqrt{1-\rho}}\right ).
\label{eqn_lambda}
\end{equation}

The expected conditional loss rate for the normally distributed
recovery rate model (4) is easily calculated as

\begin{eqnarray} \label{eqn_s}
S(X)&=&E[\max(1-R_j,0)|X]\nonumber \\
&=&\int_{-\infty}^\infty
\max(1-\mu-\sigma\sqrt{\omega}X-\sigma\sqrt{1-\omega}z,0)f_N(z)dz \nonumber \\
&=&(1-\mu-\sigma\sqrt{\omega}X)\Phi(z_c)+\sigma\sqrt{1-\omega}f_N(z_c),
\end{eqnarray}

\noindent where $f_N(z)=\frac{1}{\sqrt{2\pi}}\exp(-z^2/2)$ is the
standard normal density function and
$$z_c=\frac{1-\mu-\sigma\sqrt{\omega}X}{\sigma\sqrt{1-\omega}}.$$
Note that in D\"{u}llmann and Trapp (2004)\nocite{DuellmannTr04} it
is approximated as

$$ \label{eqn_s0}
 S(X)=E[\max(1-R_j,0)|X]\approx
 E[(1-R_j)|X]=1-\mu-\sigma\sqrt{\omega}X,
$$

\noindent assuming that probability of $R_j$ exceeding 1 is so small
that it has no material on the results. Indeed, in the case of data
studied in this paper, the specific values of $(\mu, \sigma, \omega,
X)$ are such that the relative difference between EC calculated
using the above approximation and the closed-form formula
(\ref{eqn_s}) is less than $2\%$ for all cases. In this study,
closed-form formula (\ref{eqn_s}) will be used for all relevant
calculations.

Under the framework outlined above,  $EC^\infty$  is a function of
five model parameters $\bm{\theta}=(p,\rho,\mu,\sigma,\omega)$, with
$X=\Phi^{-1}(0.001)\approx -3.09$. Obviously, an uncertainty
 in any of the parameter estimates will cause an uncertainty in the
 EC estimate. This will be discussed in Section 4.1.

\section{Likelihood}
Consider time periods $t=1,2,\ldots ,T$ (so that $T+1$ corresponds
to the next future time period), where the following data of default
and recovery for a loan portfolio of $J_t$ firms are observed:

\begin{itemize}

\item $D_t$ -- the number of defaults in time period $t$, with $d_t$ denoting the actual realization observed;
\item $\Psi_t$ -- the  default rate in time period $t$,
$\Psi_t=D_t/J_t$, with $\psi_t$ denoting the actual realization
observed;
\item $\overline{R}_t$ -- the average recovery rate in time period $t$, with $\overline{r}_t$ denoting the actual realization
observed. \end{itemize}

\noindent Denoting the individual recovery rates for $D_t$ defaulted
firms as $R_1(t),\ldots ,R_{D_t}(t)$,  the average recovery rate is
$$\overline{R}_t=\sum_{j=1}^{D_t}R_j(t)/D_t$$
and its realization is denoted as $\overline{r}_t$.

 Also, the systematic risk factor
$X$ (latent variable) corresponding to the time periods is denoted
as
$$X_1,\ldots,X_{T+1}$$
and its realization is $x_1,\ldots,x_{T+1}$.  It is assumed that
$X_1,\ldots,X_{T+1}$ are independent and all idiosyncratic risk
factors $(Z_j,Z_j^C)$ corresponding to the time periods are all
independent.

In what follows, we derive the likelihood function of the data
required for model estimation.

\subsection{Exact Likelihood Function}

 The joint density of the number of defaults and average recovery
 rate ($D_t, \overline{R}_t)$ can be calculated by integrating out the
 latent variable $X_t$ for each time period as

\begin{equation}
f(d_t, \overline{r}_t)=\int  f(\overline{r}_t | d_t, x_t)f(d_t |
x_t)f_N(x_t)dx_t. \label{eqn_intX}
\end{equation}
Here, $f_N(\cdot)$ is the standard normal density function; and the
conditional densities $f(d_t | x_t)$ and $f(\overline{r}_t | d_t,
x_t)$ are derived below.

Given $X_t=x_t$, all firms in a homogenous loan portfolio have the
same conditional default probability
$\Pr[I_j(t)=1|X_t=x_t]=\Lambda(x_t)$ evaluated in
(\ref{eqn_lambda}). Since $D_t=\sum_{j=1}^{J_t} I_j(t)$, the
conditional distribution of $D_t$ is binomial, that is

\begin{equation}
\label{eqn_bino}
 f(d_t |
x_t)=\Pr[D_t=d_t|X_t=x_t]=\binom{{J_t}}{{d_t}}
\left(\Lambda(x_t)\right)^{d_t}
\left(1-\Lambda(x_t)\right)^{J_t-d_t}.
\end{equation}

\noindent It can be well approximated by the normal distribution
$N(\mu_t, \sigma_t^2)$ with mean $\mu_t=J_t\Lambda(x_t)$ and
variance $\sigma_t^2=J_t\Lambda(x_t)(1-\Lambda(x_t))$ if both
$J_t\Lambda(x_t)$ and $J_t(1-\Lambda(x_t))$ are larger than 5. For
the data fitted in the present study we can verify that the minimum
value for $J_t\Lambda(x_t)$ is larger than 10, and the minimum value
of $J_t(1-\Lambda(x_t))$ is much larger than 10. Thus the
distribution of $D_t$ can be approximated as

\begin{equation}
\label{eqn_default_pdf}
 f(d_t | x_t)=\frac{1}{\sqrt{2\pi}\sigma_t} \exp
\left (-\frac{(d_t-\mu_t)^2}{2\sigma_t^2}\right ).
\end{equation}

Conditional on $X_t=x_t$ and $D_t=d_t$; individual recoveries
$R_1(t),\ldots,R_{d_t}(t)$
 are independent from normal distribution
$N(\mu_r, \sigma_r^2)$ with $\mu_r=\mu +\sigma\sqrt{\omega}x_t$ and
$\sigma_r=\sigma\sqrt{1-\omega}$. Thus the average $\overline{R}_t$
is from normal distribution $N(\mu_R, \sigma_R^2)$ with
$\mu_R=\mu_r$ and $\sigma_R^2=\sigma_r^2/d_t$, i.e.

\begin{equation}
\label{eqn_recovery_pdf}
 f(\overline{r}_t| d_t,
x_t)=\frac{1}{\sqrt{2\pi}\sigma_R} \exp \left
(-\frac{(\overline{r}_t-\mu_R)^2}{2\sigma_R^2}\right).
\end{equation}

\noindent If recovery distribution is different from normal, the
average  $\overline{R}_t$ can still be approximated by normal
distribution if $d_t$ is large (and variance is finite).
Substituting (\ref{eqn_default_pdf}) and (\ref{eqn_recovery_pdf})
into (\ref{eqn_intX}), the density $f(d_t, \overline{r}_t)$ can be
 computed numerically. Define random vectors of default and recovery
rate data as
$$\bm{D}=(D_1,\ldots ,D_T)\quad\mbox{and}\quad\overline{\bm{R}}=(\overline{R}_1,\ldots
,\overline{R}_T)$$ respectively. The joint likelihood function for
data $\bm{D}$ and $\overline{\bm{R}}$ is then

\begin{equation}
\ell_{\bm{D},\overline{\bm{R}}}({\bm \theta})=\prod_{t=1}^{T}f(d_t,
\overline{r}_t). \label{eqn_joint_L}
\end{equation}

\noindent This joint likelihood function can be used to estimate
parameters $\bm{\theta}$ by MLEs maximizing this likelihood; or (as
described shortly in Section 4) posterior distribution of
$\bm{\theta}$ can be calculated using Bayesian approach via MCMC.
However, the likelihood (\ref{eqn_joint_L}) involves numerical
integration in (\ref{eqn_intX}), which integrates out the latent
variables $\bm{X}=(X_1,\ldots ,X_T)$. Our numerical experiments show
that, although not impossible,  it is difficult in practice to
accurately compute integrations in (\ref{eqn_joint_L}), especially
if the likelihood is used within a numerical maximization procedure.
One of the difficulties is the frequent occurrence of numerical
under-flow in the evaluation of the integrand, even in double
precision using Gauss-Hermite quadrature.

A more straightforward and problem-free alternative is to take
Bayesian approach and treat the latent variable  ${\bm X}$ in the
same way as other parameters, and formulate the problem in terms of
the likelihood conditional on the complete state variable vector
${\bm \gamma}=(p, \rho, \mu, \sigma, \omega, X_1,\ldots ,X_T)=({\bm
\theta}, {\bm X})$. In this case the required conditional joint
density function is

\begin{equation}
f(d_t, \overline{r}_t | x_t,{\bm \theta})=f(d_t | x_t, {\bm
\theta})f(\overline{r}_t | d_t, x_t, {\bm \theta}),
\label{eqn_joint_pdf}
\end{equation}

\noindent and the joint conditional likelihood function is

\begin{equation}
\ell_{\bm{D},\overline{\bm{R}}}({\bm \gamma})=\prod_{t=1}^T f(d_t,
\overline{r}_t | x_t,{\bm \theta}). \label{eqn_joint_Lx}
\end{equation}
Here, the integration with respect to $\bm X$ is not required and
the under-flow problem in evaluating (\ref{eqn_joint_Lx}) can now be
readily overcome by the usual approach of working with the
log-likelihood function instead. Then the samples from the joint
posterior of $({\bm \theta}, {\bm X})$ can be obtained using MCMC
and taking samples of $\bm\theta$ marginally allows to get posterior
of $\bm\theta$ effectively integrating out the latent variable $\bm
X$; this will be discussed in detail in Section 4.

Note that in the case of one latent factor, the required integration
is 1d integration, see equation (\ref{eqn_intX}), and in principle
it can be done numerically using quadrature rules.  However, in the
case of $n$-latent factors, $n$-dimension integration will be
required to get the likelihood which is not practical in the case of
two or more latent factors. Under the Bayesian approach with MCMC
method, the number of latent factors is not a problem. The
likelihood required for this procedure is just a likelihood
conditional on the latent factors. The procedure will produce
posterior samples of  parameters and latent factors, and taking
samples of the required variable marginally will effectively
integrate out other variables.

\subsection{Approximate Likelihood and Closed-Form MLEs}
By considering default and recovery processes separately and
assuming  a large number of firms in the portfolio, some
approximation can be justified to simplify the evaluation of the
likelihood function (\ref{eqn_joint_L}) and its maximization
procedures to get MLEs for the model parameters. This is the
approach taken by Frye (2000b)\nocite{Frye00} and D\"{u}llmann and
Trapp (2004)\nocite{DuellmannTr04}  that follows two stages. In the
first stage, the parameters for the default process $(\rho, p)$ and
systematic factor $\bm{X}$ are estimated. Then, the parameters of
the recovery model $(\mu, \sigma, \omega)$ are evaluated in the
second stage.

~

\noindent {\bf Default process} \\
\noindent Given $X_t$ in time period $t$, the conditional default
probability $\Lambda_t=\Lambda(X_t)$ is a monotonic function of
$X_t$; see (\ref{eqn_lambda}). The density of $X_t$ is the standard
normal, thus the change of probability measure gives the density for
$\Lambda_t$ at $\Lambda_t=\lambda_t$:

\begin{equation}
f(\lambda_t |
\bm{\theta}_D)=\frac{1}{\sqrt{2\pi}}\exp\left(-\frac{x_t^2}{2}\right)\left|
\frac{dx_t}{d\lambda_t}\right|, \label{eqn_density_lambda}
\end{equation}

\noindent where $\bm{\theta}_D=(p,\rho)$ is the parameter vector for
default process and $x_t$ is the function of $\lambda_t$, the
inverse of (\ref{eqn_lambda}),

\begin{equation}
x_t=\frac{\Phi^{-1}(p)-\sqrt{1-\rho}\Phi^{-1}(\lambda_t)}{\sqrt{\rho}}.
\label{eqn_xt}
\end{equation}

\noindent Explicitly, the density of the conditional default
probability $\Lambda_t$, is then

\begin{eqnarray}
&&f(\lambda_t|\bm{\theta}_D)=\sqrt{\frac{1-\rho}{\rho}}
\nonumber\\
&&\quad\times\exp
\left(-\frac{(\Phi^{-1}(p))^2+(1-2\rho)(\Phi^{-1}(\lambda_t))^2-2\sqrt{1-\rho}\Phi^{-1}(p)
\Phi^{-1}(\lambda_t)}{2\rho}\right).
\end{eqnarray}

\noindent For time period $t$ we observe default rate $\Psi_t$ that
(in the limit $J_t\rightarrow\infty$) approaches the conditional
default probability $\Lambda_t$. Therefore, in this limit for the
observed data vector of default rate $\bm{\psi}=(\psi_1,\ldots
,\psi_T)$, the likelihood function is

\begin{equation}
\ell_D(\bm{\theta}_D)=\prod_{t=1}^T{f(\lambda_t=\psi_t |
\bm{\theta}_D)}. \label{eqn_likelihoodD}
\end{equation}

\noindent Maximizing (\ref{eqn_likelihoodD})  gives  the following
MLEs for $\rho$ and $p$:

\begin{equation}
\hat{\rho}=\frac{\sigma_{\bm{\delta}}^2}{1+\sigma_{\bm{\delta}}^2},
\label{eqn_rho}
\end{equation}

\begin{equation}
\hat{p}=\Phi\left(\frac{\overline{\delta}}{\sqrt{1+\sigma_{\bm{\delta}}^2}}\right),
\label{eqn_p}
\end{equation}

\noindent where $\overline{\delta}=\sum_{t=1}^{T}\delta_t/T$,
$\sigma_{\bm{\delta}}^2=\sum_{t=1}^{T}(\delta_t-\overline{\delta})^2/T$
and $\delta_t=\Phi^{-1}(\psi_t)$.
 The systematic
factor $X_t$ is then estimated using (\ref{eqn_xt}) with default
parameters $(p, \rho)$ replaced by MLEs as

\begin{equation}
\hat{x}_t=\frac{\Phi^{-1}(\hat{p})-\sqrt{1-\hat{\rho}}\Phi^{-1}(\psi_t)}{\sqrt{\hat{\rho}}}.
\label{eqn_xt2}
\end{equation}

~

 \noindent {\bf Recovery process}\\
\noindent As discussed in Section 3.1, given systematic factor $X_t$
and number of defaults $D_t$, the average recovery rate
$\overline{R}_t$
 is from normal
distribution $N(\mu_R,\sigma^2_R)$ with mean
$\mu_R=\mu+\sigma\sqrt{\omega}X_t$ and variance
$\sigma^2_R=\sigma^2(1-\omega)/d_t$, and the density

\begin{equation}
f(\overline{r}_t | \bm{\theta}_R,
x_t)=\sqrt{\frac{d_t}{2\pi\sigma^2(1-\omega)}}\exp\left(
-\frac{d_t(\overline{r}_t-\mu-\sigma\sqrt{\omega}x_t)^2}{2\sigma^2(1-\omega)}\right),
\label{eqn_recovery_pdf2}
\end{equation}

\noindent where $\bm{\theta}_R=(\mu, \sigma, \omega)$; also see
(\ref{eqn_recovery_pdf}). The likelihood function for $T$
observations of the average recovery rate
$\overline{\bm{r}}=(\overline{r}_1,\ldots ,\overline{r}_T)$ is then

\begin{equation} \label{eqn_likelihoodR}
\ell_{\overline{\bm{R}}}(\bm{\theta}_R,
\bm{x})=\prod_{t=1}^T{f(r_t|\bm{\theta}_R, x_t)}.
\end{equation}

\noindent D\"{u}llmann and Trapp (2004)\nocite{DuellmannTr04}
estimate $\bm{\theta}_R$ by MLEs via maximization of
(\ref{eqn_likelihoodR}) with respect to $\bm{\theta}_R$, where $x_t$
is replaced with $\hat{x}_t$ given in (\ref{eqn_xt2}). It was found
that searching numerically for the maximum likelihood of the
recovery model may provide spurious results. Thus they took a
``feasible maximum likelihood" approach that involves two steps to
estimate the recovery parameters. In the first step, the volatility
parameter $\sigma$ was estimated by the historical volatility
\begin{equation}\label{sigma_via_histvol}
\widehat{\sigma}_h=\sqrt{\frac{1}{T-1}\sum_{t=1}^{T}(\overline{r}_t-\overline{R})^2},\hspace{1cm}
\overline{R}=\frac{1}{T}\sum_{t=1}^{T}\overline{r}_t.
\end{equation}

\noindent In the second step, parameters $\mu$ and $\omega$ were
estimated conditional on $\sigma=\widehat{\sigma}_h$. It is
important to note that setting $\sigma=\widehat{\sigma}_h$ is
conceptually incorrect because this historical volatility
$\widehat{\sigma}_h$ is the volatility of the average annual
recovery rates that does not include the cross-section variability,
while model parameter $\sigma$ is the measure of the overall
recovery variability. One can easily correct this by setting
$\sigma\sqrt{\omega}=\widehat{\sigma}_h$ which is valid in the limit
of large number of defaults.

We met with similar numerical difficulties when trying to estimate
$(\mu, \sigma, \omega)$ jointly by numerical minimization of the
log-likelihood function. However, re-parameterizing with
$\sigma_1=\sigma\sqrt{\omega}$ and $\sigma_2=\sigma\sqrt{1-\omega}$,
a closed-form solution for MLEs of $(\mu, \sigma, \omega)$ can be
easily obtained. Let
$G(\bm{\theta}_R,\bm{x})=\ln(\ell_{\overline{\bm{R}}}(\bm{\theta}_R,
\bm{x}))$, then solving ${\partial G}/{\partial \mu}=0$, ${\partial
G}/{\partial \sigma_1}=0$ and   ${\partial G}/{\partial \sigma_2}=0$
gives the following closed-form MLEs

\begin{equation} \label{eqn_sigma1}
\widehat{\sigma}_1=\frac{\left(\sum_t d_t\overline{r}_tX_t
\right)\left(\sum_t d_t\right) - (\sum_t d_t\overline{r}_t )(\sum_t
d_tX_t ) }{\left(\sum_t d_tX_t^2\right)\left(\sum_t d_t\right)
-\left(\sum_t d_tX_t \right)^2 },
\end{equation}

\begin{equation} \label{eqn_mu}
\widehat{\mu}=\frac{(\sum_t d_t\overline{r}_tX_t ) - (\sum_t d_t
X_t^2 )\widehat{\sigma}_1 }{\sum_t d_tX_t},
\end{equation}

\begin{equation} \label{eqn_sigma2}
\widehat{\sigma}_2=\sqrt{\frac{1}{T}\sum_t d_t
(r_t-\widehat{\mu}-\widehat{\sigma}_1 X_t)^2 },
\end{equation}

\begin{equation} \label{eqn_omega}
\widehat{\omega}=\frac{\widehat{\sigma}_1^2}{\widehat{\sigma}_1^2+\widehat{\sigma}_2^2},
\end{equation}

\begin{equation} \label{eqn_sigma}
\widehat{\sigma}=\sqrt{\widehat{\sigma}_1^2+\widehat{\sigma}_2^2 }.
\end{equation}

~

\noindent\textbf{Remarks}
\begin{itemize}
\item Note that in Frye (2000b)\nocite{Frye00}, estimation procedure
is presented for the case when the overall fitted default rate
consists of defaults from firms with different rating grades (AAA,
AA1,\ldots, CA, C) assuming different probability of default $p$ for
each grade. Then, the probability of default $p$ for a specific
rating grade is estimated as a long-term average default rate of
firms in this grade; parameter $\rho$ is the same for all firms and
is estimated using the maximum likelihood method; and systematic
factor $x_t$ is implied. Estimation procedure for recoveries is
presented for the case when the overall fitted recovery rate
consists of recoveries from firms with different seniority classes
(senior secured, senior unsecured, senior subordinated and
subordinated) assuming different parameter $\mu$ for each seniority
class. Then $\omega$, $\sigma$ and all parameters $\mu$ are
estimated by maximum likelihood method.

\item Experienced numerical instabilities when estimating recovery
parameters using MLE are due to the fact that we fit time series of
average recoveries $\overline{R}_t$ whose variance
$\sigma^2_R=\sigma^2(1-\omega)/d_t$ will tend to zero for large
number of defaults $d_t$ causing flatness of the likelihood. Thus it
will be impossible to estimate recovery parameters in the limit of
large $d_t$ using the above described two-stage procedure. Ideally,
we need time series of individual recoveries ${R}_j(t)$ to avoid
this problem.

\item Note that in the above described two-stage procedure, systematic factor $x_t$
is estimated from defaults assuming fully diversified portfolio
(large number of borrowers and defaults) and then substituted into
the recovery process where the assumption of fully diversified
portfolio is not used. Under the valid statistical approach,
systematic factor $x_t$ should be estimated using information both
from defaults and recoveries. This can be achieved by maximizing the
proper joint likelihood (\ref{eqn_joint_L}). However, the presented
two-stage procedure is intuitively appealing and produce reasonable
estimates at  least for the case of data considered in this paper.
\end{itemize}

\section{Bayesian Inference and MCMC}
\label{sec:bayesian} Bayesian inference is a convenient approach to
jointly estimate all model parameters and latent factors, and all
relevant uncertainties in the model. It is especially useful when
data are limited and parameter uncertainty is large. In this case
Bayesian approach is superior to the maximum likelihood method that
relies on a large sample limit Gaussian approximation for the
parameter uncertainty. Under the Bayesian approach, the inference is
based on the distribution of the parameters and latent factors given
data (so-called posterior distribution). Typically, the posterior
distribution is not available in closed-form but can be easily
estimated numerically using MCMC method. In this section, we
introduce the main notation and concepts for Bayesian approach and
present MCMC algorithm. This well known material is presented in
this section for the benefit of the readers who are not familiar
with Bayesian inference and MCMC. There is a broad literature
covering Bayesian inference and its applications, for example, see
\nocite{RobS94} Robert and Smith (1994), Lee (1997),
\nocite{Berger85} Berger (1985), \nocite{Rob01} Robert (2001),
\nocite{Win03} Winkler (2003), \nocite{GelCS03}  Gelman et al
(2003), \nocite{Bol04} Bolstad (2004) and \nocite{CarL08} Carlin and
Louis (2008). In particular, recent examples of applying Bayesian
inference in operational risk and insurance modeling are found in
Shevchenko (2011)\nocite{Shevchenko2011} and Peters \textit{et al}
(2009a, 2009b)\nocite{PeShWu09}\nocite{PeShWu09b}.

\subsection{Bayesian Inference Approach}
Consider a random vector of data $\bm{Y}$ whose density for a given
vector of parameters ${\bm \theta }$ is $\pi ({\rm {\bf y}}\vert
{\bm \theta})$. In the Bayesian approach, both data and parameters
are considered to be random. A convenient interpretation is to think
that parameter is a random variable $\bm\Theta$ with some
distribution and the true value (which is deterministic but unknown)
of the parameter is a realization of this random variable. Then the
joint density of the data and parameters is
\begin{equation}
\pi ({ {\bm y}},{\bm \theta }) = \pi ({{\bm y}}\vert {\bm
\theta})\pi ({\bm \theta})=\pi ({\bm \theta}\vert {{\bm y}})\pi
({{\bm y}}),\label{JointDensityForBayesTheorem}
\end{equation}
where
\begin{itemize}
\item $\pi ({\bm \theta})$ is the density of parameters (a so-called
\emph{prior density});
\item $\pi ({\bm \theta}\vert {{\bm y}})$ is the density of
parameters given data $\bm Y=\bm y$ (a so-called \emph{posterior
density});
\item $\pi ({ {\bm y}},{\bm \theta })$ is the joint density of the
data and parameters;
\item $\pi ({{\bm y}}\vert {\bm\theta})$ is the density of the data
given parameters  $\bm\theta$. This is the same as a likelihood
function $\pi ({\rm {\bf y}}\vert {\bm \theta
})=\ell_{\bm{Y}}(\bm{\theta})$ given by (\ref{eqn_joint_L}) for the
model we study;
\item $\pi
({{\bm y}})$ is the marginal density of $\bm Y$, i.e. $\pi ({{\bm
y}})=\int {\pi ({\rm {\bf y}}\vert {\bm \theta })\pi ({\bm \theta
})d{\bm \theta }}$.
\end{itemize}

Using (\ref{JointDensityForBayesTheorem}), the well-known Bayes's
theorem says that the posterior density can be calculated as
\begin{equation}
\pi ({\bm \theta}\vert {{\bm y}})= \pi ({{\bm y}}\vert {\bm
\theta})\pi ({\bm \theta})/\pi ({{\bm y}})\propto \pi ({{\bm
y}}\vert {\bm \theta})\pi ({\bm \theta}).\label{SimpleBayesTheorem}
\end{equation}
Here $\pi ({{\bm y}})$ plays the role of a normalization constant.
Under the pure Bayesian approach, the prior $\pi(\bm{\theta})$
should be specified subjectively by the modeller. If there is no
prior knowledge and we would like to rely only on data to make
inference, then one can use noninformative priors such as constant
prior (i.e. uniform distribution).

The posterior can be used for predictive inference and
quantification of parameter uncertainty. For example, using the
posterior $\pi ({\theta}\vert {{\bm y}})$, one can easily construct
a credibility interval $[a,b]$ to contain the true value of the
parameter with probability
$$\Pr[a\le \Theta\le b]=\int_a^b
\pi ({\theta}\vert {{\bm y}})d\theta.$$ This is analogue for
confidence intervals under the frequentist approach but these
intervals are conceptually different. The bounds of the frequentist
confidence interval are considered to be random (functions of random
data) while bounds of the Bayesian credibility interval are
functions of data realization. Generally speaking, the variability
in posterior (e.g. its standard deviation) is due to finite data
size; increasing data size will decrease the standard deviation of
the posterior.

Typical point estimates of the parameter $\theta$ are the mean and
mode of the posterior density (depending on objective function)
called the Minimum Mean Square Estimator (MMSE) and the Maximum a
Posteriori (MAP) estimator respectively. It is obvious from
(\ref{SimpleBayesTheorem}) that if the prior is constant and the
parameter range includes the MLE then the mode of the posterior is
the same as MLE.

Denote the posterior mode as $\widehat{\bm\theta}^{MAP}$. If the
prior is continuous at the mode, it is illustrative to consider a
Gaussian approximation for the posterior obtained by a second-order
Taylor series expansion around $\widehat{\bm\theta}^{MAP}$,

\begin{equation}
\pi ({\bm \theta}\vert {{\bm y}})\approx \pi (\widehat{\bm
\theta}^{MAP}\vert {{\bm
y}})+\frac{1}{2}\sum_{i,j}\left.\frac{\partial^2\ln \pi ({\bm
\theta}\vert {{\bm y}})}{\partial\theta_i\partial\theta_j}
\right|_{\bm\theta=\widehat{\bm\theta}^{MAP}}
(\theta_i-\widehat\theta_i^{MAP})(\theta_j-\widehat\theta_j^{MAP}).
\end{equation}
Under this approximation, $\pi ({\bm \theta}\vert {{\bm y}})$ is a
multivariate normal with mean $\widehat{\bm\theta}^{MAP}$ and
covariance matrix calculated as the inverse of matrix
$(\mathbb{I})_{ij}=-\partial^2\ln \pi ({\bm \theta}\vert {{\bm
y}})/\partial\theta_i\partial\theta_j$ at $\bm
\theta=\widehat{\bm\theta}^{MAP}$. It is easy to see that this
matrix $\mathbb{I}$ in the case of improper constant prior is the
same as the observed information matrix often used to calculate
errors of MLEs.

Typically, for small datasets, the parameter uncertainity is large
and Gaussian approximation for the posterior cannot be used as well
as the large sample Gaussian approximation cannot be used for
maximum likelihood estimators. In this case, one has to evaluate the
posterior distribution (\ref{SimpleBayesTheorem}). The explicit
evaluation of the posterior often cannot be done in closed form and
numerical methods should be used. MCMC method is an efficient
technique to get samples from the posterior; one of the simplest
MCMC algorithms will be presented in Section
\ref{subsec:metropolis}.

In the one-factor credit risk model studied in this paper, the
systematic risk factor $\bm{X}=(X_1,\ldots,X_T)$ for the observed
data period is a latent random variable. It should be integrated out
to evaluate the likelihood $\pi ({\rm {\bf y}}\vert {\bm \theta
})=\ell_{\bm{Y}}(\bm{\theta})$ given by (\ref{eqn_joint_L}). Then,
the posterior $\pi ({\bm \theta }\vert {\rm {\bf y}})$ can be
calculated using (\ref{SimpleBayesTheorem}). The required
integration might be difficult and can be avoided by considering the
joint posterior of both $\bm\theta$ and $\bm{X}$, i.e. $\pi ({\bm
\gamma }\vert {\rm {\bf y}})$ with ${\bm \gamma}=({\bm \theta },
{\bm X})$. Given a \textit{prior} density $\pi ({\bm \gamma })$ and
a likelihood $\pi ({\rm {\bf y}}\vert {\bm \gamma
})=\ell_{\bm{Y}}(\bm{\gamma})$, the \textit{posterior} density is
just
\begin{equation}
\pi ({\bm \gamma }\vert {\rm {\bf y}}) \propto \pi ({\rm {\bf
y}}\vert {\bm \gamma })\pi ({\bm \gamma }). \label{eqn_bayes}
\end{equation}
Here, the likelihood $\pi ({\rm {\bf y}}\vert {\bm \gamma })$ is
given by (\ref{eqn_joint_Lx}) that does not involve integration;
also the prior for $X_t$ is the standard normal density. Then MCMC
can be used to get samples from the posterior $\pi ({\bm \gamma
}\vert {\rm {\bf y}})$, i.e. joint samples of model parameters ${\bm
\theta }$ and latent factor ${\bm X}$. Taking samples of ${\bm
\theta }$ marginally, we can get the posterior for model parameters
$\pi ({\bm \theta }\vert {\rm {\bf y}})$, i.e. effectively
integrating out the latent factor $\bm X$. Similarly, taking samples
of ${X_t}$ marginally, we can get the posterior for systematic
factor $\pi ({X_t }\vert {\rm {\bf y}})$. In this way, MCMC will
estimate parameters and latent variables simultaneously.

\subsection{Quantile Estimates Accounting for Parameter Uncertainty}
Bayesian methods are particularly convenient to quantify parameter
uncertainty and its impact on quantile estimate; see for example
Shevchenko (2008)\nocite{Shevchenko08a}. Under the Bayesian
approach, the full predictive density (accounting for parameter
uncertainty) of the next time period loss $L_{T + 1}$, given all
data ${\rm\bf{Y}}$ used in the estimation procedure, is
\begin{equation}
\label{FullPredPDF_eq} f_{L_{T+1}}(z\vert {\rm {\bf y}}) = \int
{f_{L_{T+1}}(z\vert {{\bm\theta }}) \pi ({{\bm\theta }}\vert {\rm
{\bf y}})d{{\bm\theta }}}.
\end{equation}

\noindent Here, it is assumed that, given ${{\bm \Theta }}$, $L_{T +
1}$ and ${\rm {\bf Y}}$ are independent. The quantile of the full
predictive density (\ref{FullPredPDF_eq}),
\begin{equation}
\label{QuantileFullPred_eq} Q_q^P =  \inf \{z:\Pr [L_{T + 1}
> z\vert {\rm {\bf Y}}] \le 1 - q\},
\end{equation}

\noindent at the level $q$, can be used as a risk measure for
capital calculations. Here, ``$P$'' in the upper script is used to
emphasize that this is a quantile of the full predictive
distribution. The procedure for
 simulating $L_{T+1}$ from (\ref{FullPredPDF_eq}) and calculating $Q_q^P$ using posterior samples of parameters
 can be described as follows.

\vspace{0.3cm} \noindent {\bf   Algorithm 2} ({\bf Quantile of full
predictive loss distribution})

\begin{enumerate}
  \item Draw a sample ${\bm\theta }$ for the parameters from the posterior
  density $\pi({\bm\theta } | {\bm y})$ (an efficient sampling technique is  MCMC, and the details of which will
  be described in Section 4.2).
  \item Given  posterior sample ${\bm\theta }$ for the parameters, simulate loss $L$ following  steps 1 to 4 in Algorithm 1.
\item Repeat the above steps 1-2 to obtain $N$ samples
of $L$.
\item Estimate $Q^P_q$ using obtained samples of $L$ in
the standard way.
\end{enumerate}

~

\noindent {\it \bf Distribution of quantile estimate}\\
\noindent Another approach under a Bayesian framework to account for
parameter uncertainty is to consider a quantile $Q_q ({{\bm \Theta
}})$ of the conditional  loss density $f(\cdot\vert {{\bm \Theta
}})$,

\begin{equation}
\label{DistrOfQuantile_eq} Q_q ({{\bm \Theta }}) =  \inf \{z:\Pr
[L_{T + 1} > z\vert {{\bm \Theta }}] \le 1 - q\}.
\end{equation}

\noindent Then, given that ${{\bm \Theta }}$ is distributed as $\pi
({{\bm \theta }}\vert {\rm {\bf y}})$, one can find the associated
distribution of $Q_{q} ({{\bm \Theta }})$ and form a predictive
interval to contain the true quantile value with some probability.
Under this approach, one can argue that the conservative estimate of
the capital  accounting for parameter uncertainty should be based on
the upper bound of the constructed predictive interval. However it
might be difficult to justify the choice of the required confidence
level for this interval; e.g. is it enough to take the 0.99
confidence level for estimating 0.999 quantile? The following
algorithm can be used to obtain the posterior distribution of
quantile $Q_{q} ({{\bm \Theta }})$.

~

\noindent {\bf Algorithm 3} ({\bf Distribution of quantile})
\begin{enumerate}
  \item Draw a sample ${\bm\theta }$ for the parameters from the posterior
  density $\pi({\bm\theta } | {\bm y})$. This can be done using MCMC
   described in Section 4.2.
  \item Compute $Q_q=Q_q({\bm \theta})$ using e.g. Algorithm 1.
\item Repeat the above steps to obtain $N$ samples
of $Q_q ({{\bm \Theta }})$.
\end{enumerate}

In practice the above procedure for
 simulating the distribution of $Q_q ({{\bm \Theta }})$
 can be time consuming,  because it
involves a long loop (Algorithm 1) inside the loop over parameter
samples. However, for some limiting cases considered below, the
inner loop (Step 2) can be approximated by a closed-form formula and
thus making the calculation of  the distribution of $Q_q ({{\bm
\Theta }})$ more affordable.

The parameter uncertainty of the quantile estimate under the
 limiting conditions can be accounted for by a simplified version of
 Algorithm 3, in which the inner loop (step 2) for computing
 quantile given parameters is replaced by $Q^{\infty}_\alpha ({{\bm \theta }})=\Lambda(X)S(X)$, with
  $X=\Phi^{-1}(1-\alpha)$, $\Lambda(X)$ given by (\ref{eqn_lambda}) and $S(X)$ by (\ref{eqn_s}).

If one would take the frequentist approach and maximum likelihood
method, then the economic capital is estimated as
$Q_q(\widehat{\bm\theta})$ where $\widehat{\bm\theta}$ is MLE. Then,
one should typically resort to a large sample limit approximation of
the parameter uncertainties by Gaussian distribution with
covariances calculated from the second derivatives of the
likelihood. Finally, the error propagation method (performing the
first-order Taylor expansion of $Q_q(\widehat{\bm\theta})$ around
${\bm\theta}$) is typically used to estimate the standard deviation
of the capital estimate $Q_q(\widehat{\bm\theta})$ via the
covariances of $\widehat{\bm\theta}$; see e.g. Shevchenko (2011,
formulas 5.14-5.16). However, the dataset for LGD model is small and
Gaussian approximation for the parameter uncertainties is certainly
not good enough (as confirmed by fitting real data in Sections 5 and
6); and thus Bayesian approach is superior.

~

\noindent \textbf{Capital loading for parameter uncertainty}\\
\noindent It is informative to calculate the extra loading for the
capital due to parameter uncertainty. This can be defined for a risk
measure $\varrho[\cdot]$ of the loss $L$ as

\begin{equation}\label{param_extra_loading}
\varrho[L]-E[\varrho[L|\bm\Theta]],
\end{equation}
where $\bm\Theta$ is a model parameter. If risk measure is the 0.999
quantile then the extra loading is
\begin{equation}\label{param_extra_loading_forVaR}
Q_{0.999}^P-E[Q_{0.999} ({{\bm \Theta }})],
\end{equation}
i.e. the difference between the quantile of the full predictive
distribution accounting for parameter uncertainty $Q_{0.999}^P$ and
posterior mean of $Q_{0.999} ({{\bm \Theta }})$. It is worth to note
that there are situations when the quantile (Value-at-Risk) is not
subadditive risk measure and there is no guarantee that the above
defined extra loading (\ref{param_extra_loading_forVaR}) is
nonnegative for all quantile levels. However, this is typically the
case for high quantiles. For a popular alternative risk measure,
expected shortfall  (which is subadditive), the extra loading
(\ref{param_extra_loading}) is guaranteed to be nonnegative. This
can be proved using Jensen's inequality; for more details, see
Denuit et al. (2005, Sections 2.6.2 and 2.6.3)\nocite{DeDhGoKa05}.

\subsection{Metropolis-Hastings Algorithm}
\label{subsec:metropolis} One of the simplest MCMC algorithms to get
samples from the posterior is Metropolis-Hastings algorithm that was
first described by Hastings (1970) as a generalization of the
Metropolis algorithm (\nocite{MetRT53}Metropolis \textit{et al}
1953). Denote the state vector ${\bm \gamma}$ at step $m$ as ${\bm
\gamma }^{(m)}$, the MCMC simulation for the present one-factor
model can be described as the follows.

\vspace{0.3cm} \noindent {\bf Algorithm 4} ({\bf Metropolis-Hastings
algorithm})

\begin{enumerate}
  \item Start with an arbitrary initial value ${\bm \gamma}^{(0)}$ for $m=0$.
  \item Generate ${\bm \gamma}^*$ from the proposal density $q({\bm \gamma}^* | {\bm
  \gamma}^{(m)})$.
\item Compute acceptance probability
$$\alpha ({\bm \gamma }^{(m)},{\bm \gamma }^ * ) = \min \left\{
{1,\;\frac{\pi({\bm \gamma }^ * | \bm{y})q({\bm \gamma}^{(m)}
 \vert {\bm \gamma }^{*})}{\pi({\bm \gamma }^{(m)}| \bm{y})q({\bm \gamma }^{*} \vert {\bm \gamma }^{(m)} )}}
 \right\}.$$
 \item Draw $u\sim U(0,1)$ (the uniform distribution), and let ${\bm \gamma}^{(m+1)}={\bm \gamma}^*$ if $u<\alpha ({\bm \gamma
}^{(m)},{\bm \gamma }^ *)$, otherwise ${\bm \gamma }^{(m+1)}={\bm
\gamma}^{(m)}$.
\item Repeat Steps 2 to 4 to obtain posterior samples for state
variable vector ${\bm \gamma}$ (collecting after burn-in period).
\end{enumerate}

From Bayes theorem (\ref{eqn_bayes}), our target distribution is the
posterior $$ \pi ({\bm \gamma}\vert {\rm {\bf y}}) \propto \pi ({\rm
{\bf y}}\vert {\bm \gamma })\pi ({\bm \gamma }),$$ i.e. it is
proportional to the product of the prior $\pi ({\bm \gamma })$ and
the model likelihood $\pi ({\rm {\bf y}}\vert {\bm \gamma
})=\ell_{\bm{D},\overline{\bm{R}}}({\bm \gamma})$ given by
(\ref{eqn_joint_Lx}).

 The single-component Metropolis-Hastings is
often efficient in practice, where  the state variable ${\bm \gamma
}$
 is partitioned into components ${\bm \gamma } = (\gamma_1
,\gamma _2 ,\dots,\gamma _n )$, which are updated one by one or
block by block. This was the framework for MCMC originally proposed
by Metropolis \textit{et al} (1953), and is adapted in this study.
Specifically, in our implementation, components $(\gamma_1 ,\gamma
_2 ,\dots,\gamma _n )$ correspond to $(p,\rho,\mu,\sigma,\omega,
X_1,\ldots,X_T)$. Other alternative MCMC methods also exist, e.g.
the univariate slice sampler utilized by Peters \textit{et al}
(2009b) for estimating model parameters and latent factors in the
context of operational risk model.

\vspace{0.5cm}
\noindent \textbf{\textit{Prior distributions}}\\
 \noindent In MCMC simulations, it is computationally more efficient to work with
 parameter $\beta=\Phi^{-1}(p)$ than directly with parameter $p$, avoiding unnecessary evaluations of $\Phi^{-1}(\cdot)$.
 In all MCMC simulation runs, we assume a uniform prior for all
parameters $(\beta,\rho,\mu,\sigma,\omega)$. The prior for latent
variable $X_t$ is the standard normal distribution and $X_1,\ldots
,X_T$ are independent. The only subjective judgement we bring to the
prior is the lower and upper bounds of the parameter values. The
range of the parameter value should be sufficiently large to allow
the posterior to be implied mainly by the observed data. In our
calculations we assume the following bounds
$$\beta\in (-10,10),\quad\rho\in (0,1),\quad \mu\in(0,1),\quad\sigma\in
(0.01,  1.0),\quad\omega\in (0, 1).$$
 That is, all parameters have
lower and upper bounds either corresponding to the full support of
the parameter domain or covering a sufficiently wide range. For
instance, the bounds  $\beta\in (-10,10)$  correspond to virtually
the full range ($0\%$ to $100\%$) of the probability of default. We
checked that
 an increase in bounds for any parameter did not lead to material difference in results.

~

The starting value of Markov chain for the $k$th component is set to
a uniform random number drawn independently from the support
$(a_k,b_k)$. For components corresponding to latent variables $X_t$,
we use support $(-5,5)$. In the single-component Metropolis-Hastings
algorithm, we adopt a truncated Gaussian distribution as the
symmetric random walk proposal density (both for parameters
$\bm\theta$ and latent variables $\bm X$). In addition, the Gaussian
density was truncated below $a_k$ and above $b_k $ to ensure each
proposal is drawn within the support of corresponding component.
Specifically, for the $k^{th}$ component at chain step $m$, the
proposal density is

\begin{equation}
q_k (\gamma ^
* \vert \gamma_k^{(m)})
 = \frac{f_N (\gamma ^
* ; \gamma _k^{(m)} , \sigma_k^{RW} )}{F_N (b_k; \gamma _k^{(m)} , \sigma_k^{RW} ) - F_N (a_k ;
\gamma_k^{(m)} , \sigma_k^{RW} )},
\end{equation}

\noindent where $f_N (\cdot; \gamma_k^{(m)} ,\sigma_k^{RW} )$ and
$F_N (\cdot; \gamma_k^{(m)} ,\sigma_k^{RW} )$ are the normal density
and distribution functions respectively, with $\gamma_k^{(m)}$ as
the mean and $\sigma_k^{RW} $ as the standard deviation. For each
component the mean of the Gaussian density was set to the current
state and the variance was pre-tuned and adjusted so as to allow the
acceptance rate to stay at or close to the optimal level. For
$d$-dimensional target distributions with i.i.d. components, the
asymptotic optimal acceptance rate was found to be 0.234
\nocite{GelGR97} (Gelman \textit{et al} 1997, \nocite{RoRo01}
Roberts and Rosenthal 2001). In pre-tuning the variances for all the
components we set 0.234 as the target acceptance rate. The above
procedure is exactly the same as in Shevchenko and Temnov
(2009)\nocite{ShTem09} or Peters \textit{et al}
(2009a)\nocite{PeShWu09}.

The MCMC run consists of three stages. In the first stage we tune
and adjust the proposal standard deviation $\sigma_k^{RW}$ to
achieve optimal acceptance rate for each component. The second stage
is the ``burn-in'' stage and samples from this period are discarded.
The last stage is the posterior sampling stage, where the Markov
chian is considered to have converged to the stationary target
distribution. Unless stated otherwise, the MCMC was performed for a
length of $N_b = 20,000$ as the ``burn-in'' period, we then let the
chain run for an additional length of $N = 100,000$ to generate the
posterior samples. Each iteration contains a complete update of all
components.

\section{Results for 1982-1999 Dataset}
\label{sec:mylabel1}

In this section we present MCMC and MLE results based on global
corporate default and bond recovery data (presented in Table
\ref{tab1}) covering the period 1982-1999, the same dataset as
analyzed in D\"{u}llmann and Trapp (2004)\nocite{DuellmannTr04}
where the reader can find a very detailed description of the raw
data and their pre-processing. The original data source is
Standard\&Poor's Credit Pro database
 (see also Standard\&Poor's 2003\nocite{SandP03}). This dataset
contains annual default rate and recovery rate observations for 18
years ($T=18$), from January 1982 to December 1999. The recovery
rates are measured either by market prices at default or prices at
emergence from default. It was observed that the estimates of the
expected recovery rates are $9\%$ - $26\%$ higher for prices at
emergence than for market prices at default. In this study we use
the definition with market prices at default.

For simplicity (just to illustrate the Bayesian approach and MCMC
method), we fit homogeneous portfolio model to the overall default
and recovery data, i.e. for all ratings and seniorities. One would
expect a better accuracy from fitting specific rating and/or
seniority buckets where homogeneous portfolio assumption is more
appropriate.

\subsection{Numerical Validation of MCMC Implementation}
Using MCMC, we get samples of model parameters $\bm{\theta}$ and
latent factor $\bm{X}$ from the posterior
$\pi(\bm{\theta},\bm{X}|\bm{Y})$ for a given dataset $\bm Y$. We
first validated our MCMC algorithm by simulating data from default
and recovery models assuming some realistic parameter values, and
performing MCMC on the simulated data. The  posterior mean should
approach the assumed parameter values used in the simulation when
the sample data size increases. Having satisfied with this
validation, we then proceed to confirm the closed-form solution
 of MLE  given in Section 3.2.

As discussed previously, the maximum likelihood procedure involves
separate stages for the default and recovery processes, and
closed-form solutions can be found for both processes. To confirm
MLE results with our MCMC simulations, we follow the same two steps.
Note that in the case of uniform prior, the posterior mode should be
the same as MLE. We first performed MCMC using the default
probability likelihood (\ref{eqn_likelihoodD}). The posterior mode
for $\rho$ and $p$ indeed agrees with those obtained using ML. Then
in the second stage we perform MCMC with the likelihood function for
recovery (\ref{eqn_likelihoodR}), conditioning on $\rho$ and $p$.
Again, the posterior modes for $\mu$, $\sigma$ and  $\omega$ agree
with the MLEs. These closed-form MLE results and the corresponding
values for the the stressed PD, the stressed LGD and corresponding
economic capital EC (i.e. calculated with $X=\Phi^{-1}(0.001)$, see
Section 2.3.2 for definitions of these quantities) are shown in
Table \ref{tab2}.

Note that, as expected, our MLE results for 1982-1999 dataset, are
the same as in D\"{u}llmann and Trapp (2004)\nocite{DuellmannTr04}
for default parameters $\widehat{p}=0.0123$ and
$\widehat{\rho}=0.0406$ but very different for recovery parameters.
This is because D\"{u}llmann and Trapp (2004)\nocite{DuellmannTr04}
estimate $\sigma$  by historical volatility
$\widehat{\sigma}_h=0.0845$ calculated using
(\ref{sigma_via_histvol}), which is not a valid approximation, and
then estimate $\widehat{\mu}= 0.438$ and $\widehat{\omega}=0.0998$.

 Thereafter, unless otherwise stated, MCMC results correspond to
the full conditional joint likelihood function (\ref{eqn_joint_Lx}),
i.e. without the approximations discussed in Section 3.2.

\subsection{Posterior Distributions}

After validating our numerical algorithm, the full MCMC simulation
was run, using likelihood function (\ref{eqn_joint_Lx}) for default
probability parameters $(p, \rho)$ and recovery parameters $(\mu,
\sigma, \omega)$, treating latent variables $\bm{X}=(X_1,\ldots
,X_T)$ as parameters. That is, MCMC gives samples from posterior
distributions for parameters $\bm \theta$ and latent factor ${\bm
X}$.

\begin{figure}[!htbp]
 \centerline{\includegraphics[scale=1.0]{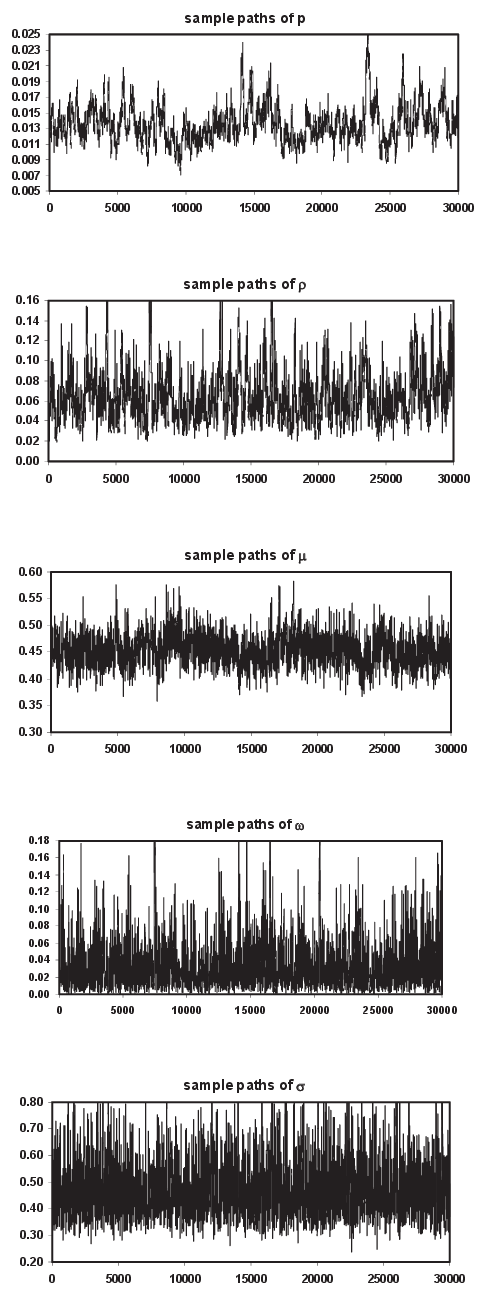}}
\caption{Markov chain Monte Carlo sample paths for parameters $(p,
\rho, \mu, \omega, \sigma)$ for the case of the 1982-1999 dataset.}
\label{fig1}
\end{figure}

\begin{figure}[!htbp]
\centerline{\includegraphics[scale=0.9]{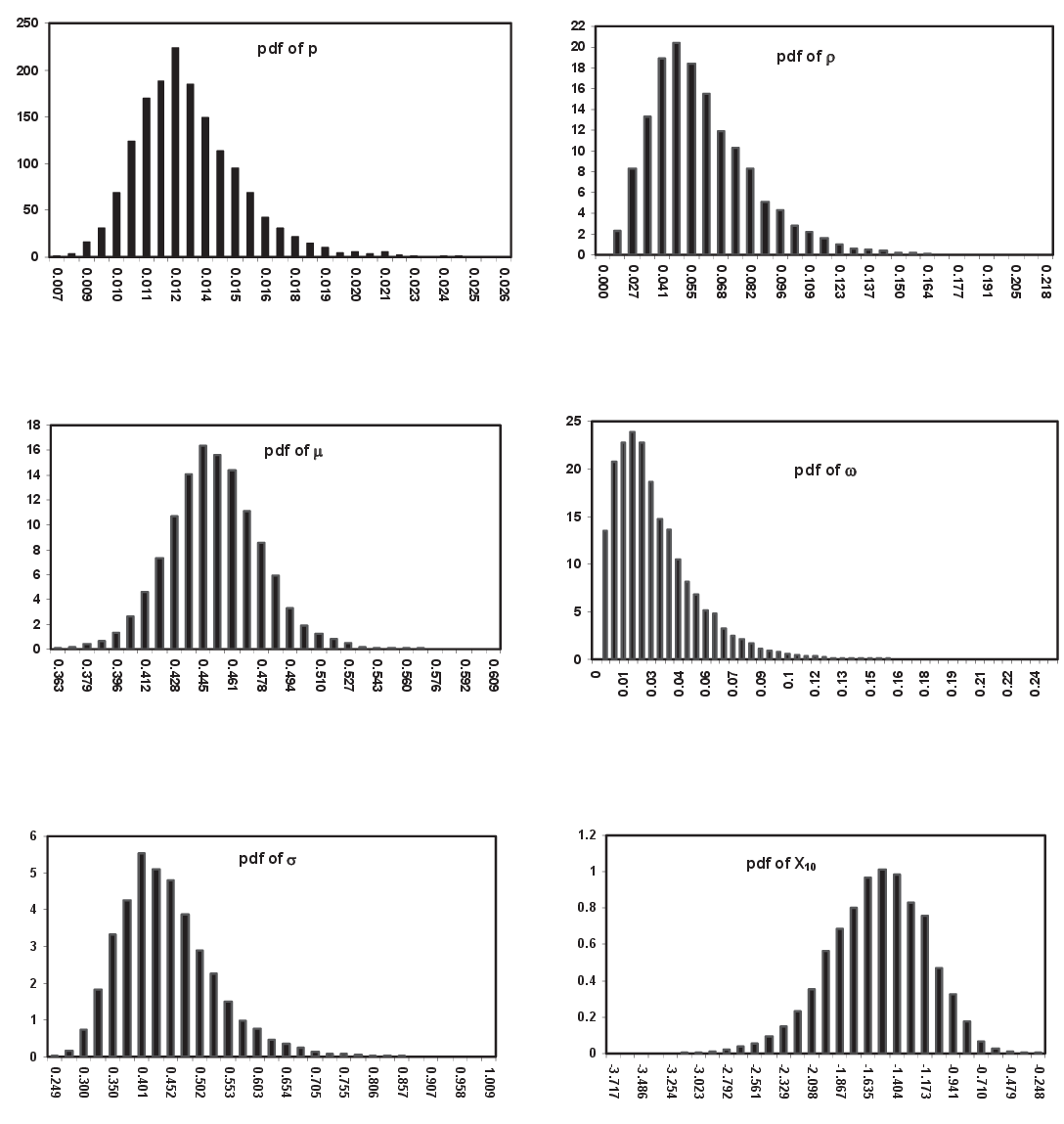}} \caption{Posterior
density functions for the model parameters $(p, \rho, \mu, \sigma,
\omega)$ and systematic factor $X_{10}$ (corresponding to the year
of 1991), computed from MCMC samples for the 1982-1999 dataset.}
\label{fig2}
\end{figure}

 The posterior sample
paths (after the burn-in) for parameters $(p,\rho,\mu, \omega,
\sigma)$ are shown in Figure \ref{fig1}. All paths reveal well-mixed
MCMC samples indicative of stationary distributions, as expected for
a convergent MCMC simulation. Figure \ref{fig2} shows the posterior
density functions estimated from the posterior samples for
parameters $(p,\rho,\mu, \omega, \sigma)$ and one of the 18 latent
variables $X_{10}$. Clearly, the densities show some positive
skewness for all parameters and negative skewness for $X_{10}$.
Table \ref{tab3} shows the summary statistics of all five parameters
computed from the posterior samples -- the mode, mean, standard
deviation (stdev), skewness and kurtosis. To quantify uncertainty in
a simple manner, the coefficient of variation (CV), defined as the
ratio of standard deviation to the mean, is also shown in Table
\ref{tab3}. Consistent with Figure \ref{fig2}, we see significant
positive skewness in most parameters; kurtosis values of  some
parameters are significantly larger than three (kurtosis of a normal
density). This also indicates that Gaussian approximation for
parameter uncertainties (typically used under the frequentist
maximum likelihood method) is not appropriate.

\subsection{Impact of Parameter Uncertainty on Quantile Estimate}
Comparison between Tables \ref{tab2} and \ref{tab3} shows that the
closed-form MLEs for all parameters are within one standard
deviation from the posterior mean.

\begin{figure}[!htbp]
\centerline{\includegraphics[scale=0.7]{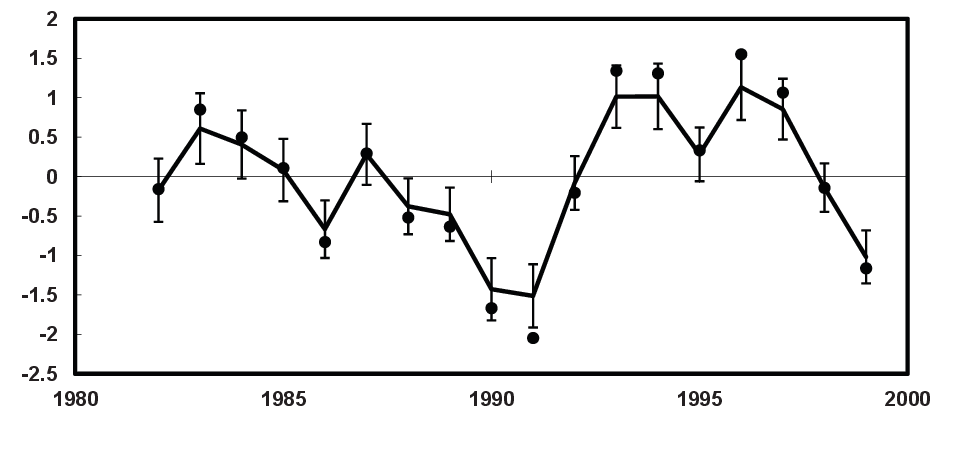}} \caption{Posterior
mean (given data) for the systematic factor $X_t, t=1,\ldots ,18$
(solid line) in comparison with the maximum likelihood point
estimates (dots), corresponding to the 1982-1999 dataset. Error bars
correspond to posterior standard deviation of $X_t$.} \label{fig3}
\end{figure}

The comparison for the 18 latent variables of systematic factor (for
18 years of data between 1982-1999) $X_t, t=1,\ldots ,18$ is shown
in Figure \ref{fig3}. Here, $X_t$ implied by MLEs (\ref{eqn_xt})
agrees well with the corresponding posterior sample mean -- only one
point is more than one standard deviation away from the posterior
mean.

A very large difference in model parameters does not always imply a
large difference in model predictions.  The predictions on the
stressed PD, LGD and EC are shown in Table \ref{tab2} for MLE and
Table \ref{tab4} for MCMC. The quantiles in Table \ref{tab4} were
obtained from Algorithm 3. For a comparison between point estimates,
the point estimates for PD, LGD and EC using the posterior mean
$\hat{\bm\theta}^{\tiny{\mbox{MMSE}}}=E[{\bm\theta}|{\bm Y}]$
(instead of MLEs) are
$$PD(\hat{\bm\theta}^{\tiny{\mbox{MMSE}}})=0.0682,\quad
LGD(\hat{\bm\theta}^{\tiny{\mbox{MMSE}}})=0.776 \quad\mbox{and}\quad
Q^{\infty }_{0.999}(\hat{\bm\theta}^{\tiny{\mbox{MMSE}}})=0.054.$$

\begin{figure}[!htbp]
\centerline{\includegraphics[scale=1]{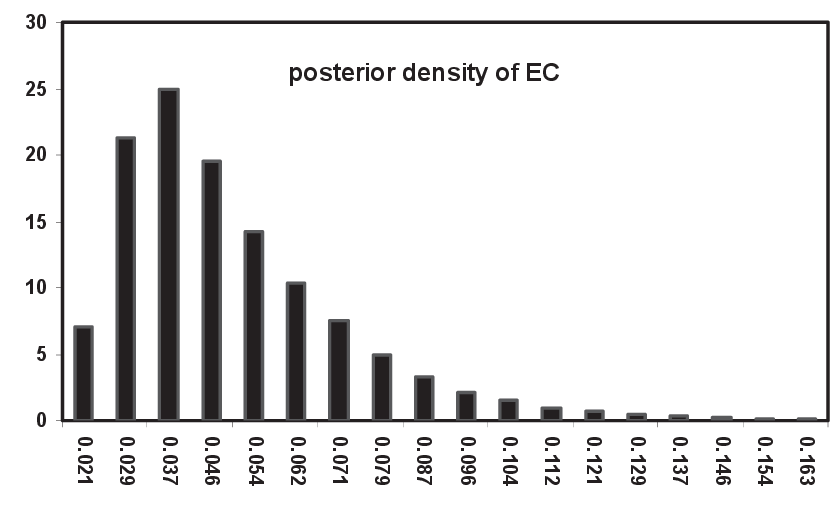}} \caption{Posterior
density of economic capital $EC^\infty=Q^{\infty}_{0.999} ({\bm
\Theta })$ computed from MCMC samples using Algorithm 3 for the
1982-1999 dataset.} \label{fig4}
\end{figure}

The posterior density of EC is shown in Figure \ref{fig4}. Evidently
the distribution is positively skewed. Comparison of Table
\ref{tab2} and \ref{tab4} shows that for the EC as defined in
(\ref{eqn_ec}), the MLE point estimate (in closed form)  is $58\%$
lower than the posterior mean from MCMC, $41\%$ lower than the
posterior median and about $90\%$ lower than the 0.75 quantile. The
MLE is within one standard deviation from the posterior mean.
However, note that the uncertainty (due to small data size) is very
large, CV is about 42\%; also note a large difference between the
0.75 and the 0.25 quantiles of EC posterior. Given that posterior
density of EC is skewed, CV is not a good measure of parameter
uncertainty and it is better to use posterior quantiles for this
purpose. The underestimation of EC by the MLE in comparison with
Bayesian posterior estimates is quite significant, and this is the
consequence of large parameter uncertainty, and large skeweness in
posterior of EC. The latter also indicates that the use of the error
propagation method based on the first-order Taylor expansion to
estimate the error in EC via the errors in parameters (typically
used under the frequentist approach) would not be appropriate.

\subsection{Quantile Estimate via Full Predictive Loss Distribution}
Table \ref{tab5} shows the 0.999 quantile ${Q}^P_{0.999}$ of the
full predictive loss density $f_{L_{T+1}}(\cdot|\bm{y})$ (the
density of loss given data only, where parameters are integrated
out; see Section 4.1), in the case of portfolios with a different
number of borrowers $J$ assuming equal weights $w_1=\cdots=w_J=1/J$.
The highest number $J=5000$ is close to the actual number of firms
in the last year of the 18 year dataset. The qauntiles
${Q}^P_{0.999}$ in Table \ref{tab5} were computed using Algorithm 2.
At $J=\infty$, instead of using Step 2 of Algorithm 2, the loss $L$
is calculated using formula (\ref{eqn_infL2}) for the limiting case
of a large portfolio. Clearly from Table \ref{tab5}, the quantile
$Q^P_{0.999}$ decreases with the number of firms $J$, reaching a
limiting value at $J=\infty$. The smaller quantile for the loss
distribution of a larger portfolio is a diversification effect.
 For instance for the period 1982-1999,  at $J=500$ the full
predictive quantile $Q^P_{0.999}$ is about $30\%$ lower than the
case at
 $J=50$; the quantile $Q^P_{0.999}$ for the limiting case $J=\infty$ is about $4\%$ lower than for the $J=500$ case.

\begin{figure}[!htbp]
\centerline{\includegraphics[scale=1]{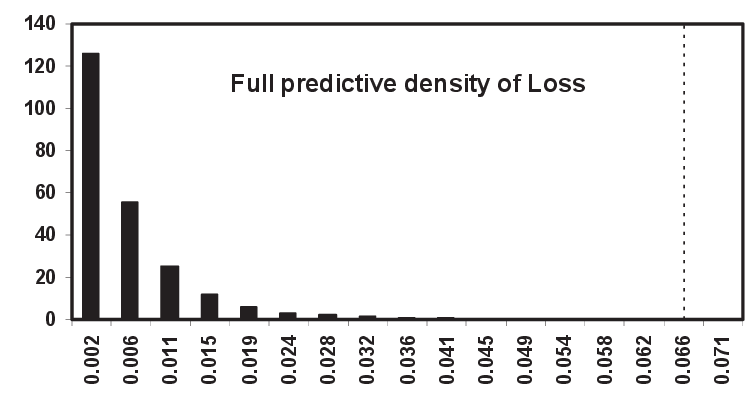}} \caption{The full
predictive  density (accounting for parameter uncertainty) of the
total loss $L^\infty_{T+1}$ computed from MCMC samples using
Algorithm 2 for the 1982-1999 dataset. Dashed line indicates the
0.999 quantile, $Q^P_{0.999}$.} \label{fig5}
\end{figure}

The posterior density of the full predictive distribution for
$L^\infty$  is shown in Figure \ref{fig5}. The quantile of full
predictive distribution ${Q}^P_{0.999}$ at $J=\infty$ is more than
twice as large as the $EC^{\infty}$ estimated by the approximate MLE
(shown in Table \ref{tab2}), and it is also $30\%$ larger than the
posterior mean of $Q^{\infty}_{0.999} ({\bm \Theta })$ (shown in
Table \ref{tab4}). This illustrates that parameter uncertainty is
very significant in determining economic capital in the one-factor
credit risk model studied here, which is not surprising given a
small dataset of annual defaults and recoveries over 18 years. Also,
this shows that the use of MLE may lead to a very significant
underestimation in EC.

To account for parameter uncertainty (due to finite sample size), we
suggest that EC should be measured as the quantile ${Q}^P_{q}$ of
the full predictive loss distribution rather than some point
estimates based on MLEs or characteristics of the posterior for
$Q^{\infty}_q ({\bm \Theta })$. The extra loading in EC due to
parameter uncertainty can be defined by
(\ref{param_extra_loading_forVaR}), i.e. the difference between
${Q}^P_{0.999}$ and posterior mean of $Q^{\infty}_{0.999} ({\bm
\Theta })$.

\section{Results for 1982-2010 Dataset}
\label{sec:sec6} In this section we show MCMC and MLE results based
on the default and recovery data (presented in Table \ref{tab1})
covering the 1982-2010 period, in which the worst financial crisis
since the Great Depression of the 1930 occurred. The historical data
for corporate default and recovery rates were taken from Moody's
(2011)\nocite{Moody2011}. The year 1982 is the earliest year for the
recovery data provided in the Moody's report. Also note that Moody's
and Standard\&Poor's data for 1982-1999 period are almost the same.

\begin{figure}[!htbp]
\centerline{\includegraphics[scale=0.9]{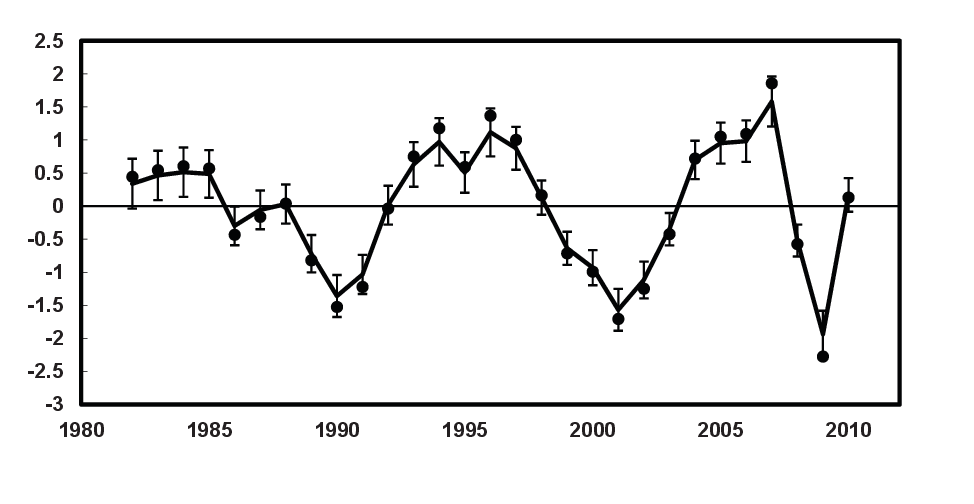}}
\caption{Posterior mean (given data) of systematic factor $X_t,
t=1,\ldots ,29$ (solid line) in comparison with the maximum
likelihood point estimates (dots), corresponding to the 1982-2010
dataset. Error bars correspond to posterior standard deviation of
$X_t$.} \label{fig6}
\end{figure}

The longer time period 1982-2010 has 11 extra recent years compared
with the earlier period 1982-1999 considered in Section 5. Using
this longer dataset, the closed-form MLE results for default and
recovery parameters and the corresponding values for the the
stressed probability of default PD, the stressed LGD and the
economic capital EC are shown in Table \ref{tab2}. The results for
1982-2010 dataset certainly have higher probability of default PD,
higher loss in terms of LGD and higher economic capital EC when
compared to the results obtained from 1982-1999 dataset. Obviously,
this is due to the global financial crisis occurred in recent years.
The systematic factor $X_t$ for 2009 was found to be -2.27, which
corresponds to approximately $99\%$ quantile level of  the limiting
loss distribution of the diversified portfolio. This maximum
negative systematic factor for  2009 is the consequence of the
disastrous 2008 when the bankruptcy of Lehman Brothers occurred (the
largest bankruptcy filing in U.S. history). The comparison for the
29 latent variables $X_t, t=1,\ldots ,29$ (corresponding to 29 years
of 1982-2010 dataset)  is shown in Figure \ref{fig6}. The systematic
factor $X_t$ implied by MLE parameter values (\ref{eqn_xt}) again
agrees well with the posterior sample mean of MCMC; all maximum
likelihood point estimates are within one standard deviation from
the posterior mean.

The summary statistics (mode, mean, standard deviation, skewness,
kurtosis and the coefficient of variation) of all five model
parameters computed from the posterior samples for the 1982-2010
dataset are shown in Table \ref{tab3}.
 Similar to the period of 1982-1999, we see significant
positive skewness in most parameters. In addition, the kurtosis
values of some parameters are significantly higher than kurtosis of
a normal density.

 Results for the 1982-2010 dataset also
 show that the closed-form MLEs for all parameters are
within one standard deviation from the posterior mean (see Tables
\ref{tab2} and \ref{tab3}). Again the closed-form MLE solution for
$\sigma$ is close to the posterior mean for the 1982-2010 dataset.
 The MCMC predictions on stressed PD, LGD and EC for the 1982-2010 dataset
 are also shown in  Table
\ref{tab4}. Comparison  shows that  the closed-form MLE for EC is
$35\%$ lower than the posterior mean, $24\%$ lower than the
posterior median and more than $50\%$ lower than the 0.75 quantile
of the posterior for EC. Similar to the period 1982-1999, the
uncertainty in the posterior of EC is large, CV is about 34.5\%,
even though the sample data size has increased from 18 years to 29
years. Nevertheless, the increased data size has resulted in a
reduction in uncertainty, as is evident in the reduction of CV. The
ratio of CV (standard deviation normalized by the mean) of EC for
the two time periods is $0.423/0.345\approx 1.23$ (see Table
\ref{tab4}), while the square-root ratio of the data sizes for the
corresponding time periods is $\sqrt{29/18} \approx 1.27$. This
reduction in uncertainty approximately proportional to the square
root of the data size is typically observed in statistical models,
though generally speaking is valid in the limit of large data size.

 The 0.999 quantile ${Q}^P_{0.999}$ of the
full predictive loss density $f_{L_{T+1}}(\cdot|\bm{y})$ for the
time period 1982-2010 for several portfolios with different number
of borrowers $J$ is shown in Table \ref{tab5}. Similar to the time
period 1982-1999, the
 diversification effect when increasing the number of firms from a small base is
 evident.
The quantile ${Q}^P_{0.999}$  decreases with $J$ reaching a limiting
value at $J=\infty$.
 As shown in Table \ref{tab5} for the period 1982-2010,  $Q^P_{0.999}$ at $J=500$
 is about $25\%$ lower than the case at $J=50$; and
for $J=5000$ is virtually the same as for the limiting case
$J=\infty$. Also, note that ${Q}^P_{0.999}$ at $J=\infty$ is about
50\% larger than corresponding MLE in Table \ref{tab2}; and about
15\% larger than the posterior mean of $Q^{\infty}_{0.999} ({\bm
\Theta })$ in Table \ref{tab4} (also see formula
(\ref{param_extra_loading_forVaR})), which is a significant
reduction when compared to the results for $1982-1999$ dataset. The
15\% impact of parameter uncertainty on the 0.999 quantile of the
loss distribution gives indication that $1982-2010$ dataset is long
enough for the use of the calibrated LGD model. In comparison, it
would be difficult to justify the use of the model with the
$1982-1999$ dataset where the impact of parameter uncertainty is too
large.

\section{Conclusion}
This paper presents a methodology of estimating the default and
recovery model parameters and latent systematic risk factors in the
well known LGD model via Bayesian approach and Markov chain Monte
Carlo method. Under this approach, the uncertainty in parameters and
model predictions is quantified using the posterior distribution
obtained from the prior and data likelihood. Moreover, it allows an
easy calculation of the full predictive loss density
$f_{L_{T+1}}(\cdot|\bm y)$ accounting for parameter uncertainty as
described in Section 4.1; then the economic capital can be based on
the high quantile of this distribution $Q_q^{P}$.

Given small datasets typically used to fit the model, the parameter
uncertainty is large and the posterior is very different from the
normal distribution indicating that Gaussian approximation for
parameter uncertainties (typically used under the frequentist
maximum likelihood approach assuming large sample limit) is not
appropriate. As an illustration, using Moody's and Standard\&Poor's
data for the annual corporate default and recovery rates, we
calibrated the model and quantified the impact of parameter
uncertainty on economic capital as if this dataset would correspond
to the dataset of the real bank portfolio. The posterior mean of
economic capital $Q_{0.999}^{\infty}(\bm\Theta)$ is 35\% higher than
corresponding MLE estimate for the longest 1982-2010 dataset, and
58\% higher for the 1982-1999 dataset. In addition, the 0.999
quantile of the full predictive distribution $Q_{0.999}^{P}$ is more
than twice as large as the MLE estimate of EC for 1982-1999 dataset
and about 50\% larger for 1982-2010 dataset. This strongly indicates
that it is dangerous to use the MLE estimate for EC. The impact of
parameter uncertainty on the quantile, see formula
(\ref{param_extra_loading_forVaR}), quantified as the relative
difference between $Q_{0.999}^{P}$ and posterior mean of
$Q_{0.999}^{\infty}(\bm\Theta)$, is about 30\% for 1982-1999 dataset
and 15\% for 1982-2010 dataset. These results demonstrate that the
extra capital to cover parameter uncertainty can be significant and
should not be disregarded by practitioners developing LGD models.

At this stage, datasets of default and recovery time series for bank
loans are not available and thus the considered LGD model cannot be
used for direct calculations of capital against real credit risk
portfolios in banks. At the moment, PDs and LGDs are estimated from
balance sheet information. Therefore our numerical results for EC
and impact of parameter uncertainty on EC should be considered as an
illustration of the method only.

The main objective of the paper is to demonstrate how the Bayesian
approach and MCMC method can be used to estimate LGD model and
related quantities. For simplicity we considered the case of
homogeneous portfolio. It is not difficult to extend the approach
and algorithm to deal with non-homogeneous portfolios and more than
one latent factor. Macroeconomic factors such as GDP can be
incorporated into the model similar to R\"osch and Scheule
(2005)\nocite{Scheule05}; also it should be worth to consider mean
reversion in the systematic factor.

\section{Acknowledgement}
\label{sec:acknowledgementodel} We would like to thank the anonymous
referee for critical and constructive comments,
 Peter Thomson for useful discussions, and the credit risk
quantitative team of Commonwealth Bank of Australia for influencing
our knowledge of this subject.

\bibliography{bibliography}
\bibliographystyle{acm}

\begin{table}[htbp]
\begin{center}
\caption{Global corporate default and recovery annual rates from
Moody's (2011). Data in brackets for 1982-1999 are Standard\&Poor's
data used in D\"{u}llmann and Trapp (2004).}
\begin{tabular}{ccccc}
\toprule year &  recovery rate & default rate &  no. defaults &  no. firms\\
 \midrule
1982  &  0.353 (0.358)  & 0.01036 (0.0119) &   13 (18)   & 1255 (1513) \\
1983  &  0.445 (0.4925) & 0.00967 (0.0068) &   13 (11)   & 1344 (1618)\\
1984  &  0.455 (0.5331) & 0.00927 (0.0083) &   13 (13)   & 1402 (1566)\\
1985  &  0.436 (0.447)  & 0.00950 (0.0103) &   15 (18)   & 1579 (1748)\\
1986  &  0.474 (0.3665) & 0.01855 (0.0169) &   33 (32)   & 1779 (1893)\\
1987  &  0.513 (0.5399) & 0.01558 (0.0093) &   31 (19)   & 1990 (2043)\\
1988  &  0.388 (0.4455) & 0.01365 (0.0144) &   29 (31)   & 2125 (2153)\\
1989  &  0.323 (0.4367) & 0.02361 (0.0153) &   52 (39)   & 2202 (2549)\\
1990  &  0.255 (0.2682) & 0.03588 (0.0256) &   82 (66)   & 2285 (2578)\\
1991  &  0.355 (0.4702) & 0.03009 (0.0306) &   66 (89)   & 2193 (2908)\\
1992  &  0.459 (0.5388) & 0.01434 (0.0122) &   31 (33)   & 2162 (2705)\\
1993  &  0.431 (0.502)  & 0.00836 (0.0051) &   19 (23)   & 2273 (4510)\\
1994  &  0.456 (0.5609) & 0.00614 (0.0052) &   16 (18)   & 2606 (3462)\\
1995  &  0.433 (0.4988) & 0.00935 (0.0091) &   27 (33)   & 2888 (3626)\\
1996  &  0.415 (0.4534) & 0.00533 (0.0045) &   17 (20)   & 3189 (4444)\\
1997  &  0.488 (0.564)  & 0.00698 (0.006)  &   25 (24)   & 3582 (4000)\\
1998  &  0.383 (0.415)  & 0.01255 (0.0118) &   51 (56)   & 4064 (4746)\\
1999  &  0.338 (0.3207) & 0.02214 (0.02)   &   100 (107) & 4517 (5350)\\
2000  &  0.253          & 0.02622          &   124       & 4729 \\
2001  &  0.216          & 0.03978          &   187       & 4701 \\
2002  &  0.297          & 0.03059          &   141       & 4609 \\
2003  &  0.404          & 0.01844          &   82        & 4447 \\
2004  &  0.585          & 0.00855          &   38        & 4444 \\
2005  &  0.560          & 0.00674          &   31        & 4599 \\
2006  &  0.550          & 0.00654          &   31        & 4740 \\
2007  &  0.547          & 0.00367          &   18        & 4905 \\
2008  &  0.339          & 0.02028          &   103       & 5079 \\
2009  &  0.339          & 0.05422          &   265       & 4887 \\
2010  &  0.500          & 0.01283          &   57        & 4443 \\
 \bottomrule
\end{tabular}
\label{tab1}
\end{center}
\end{table}

\newpage

\begin{table}[!htbp]
\begin{center}
\caption{Maximum likelihood estimates of the model parameters  and
corresponding estimates of stressed PD, LGD and EC using approximate
likelihood function (\ref{eqn_likelihoodD}) for default and
(\ref{eqn_likelihoodR}) for recovery data.  Here, EC is estimated as
$Q^{\infty}_{0.999}(\hat{\bm \theta}^{MLE})$ in (\ref{eqn_ec}) with
$\hat{\bm \theta}^{MLE}$ is the maximum likelihood point estimate
for $\bm\theta$.}
\begin{tabular*}
{1.0\textwidth}{ccccccccc}
\toprule  Time period  & $p$ & $\rho$ & $\mu$ & $\sigma$ & $\omega$ & PD & LGD & EC \\
 \midrule
 1982-1999  & 0.0123 & 0.0406 & 0.450 & 0.445 & 0.0118 & 0.0488 & 0.710 & 0.0346 \\
 1982-2010 & 0.0167 & 0.0635 & 0.411 & 0.499 & 0.0192 & 0.0819 & 0.813 & 0.0666 \\
 \bottomrule
\end{tabular*}
\label{tab2}
\end{center}
\end{table}

\begin{table}[!htbp]
\begin{center}
\caption{Summary statistics of the model parameters $(p, \rho, \mu,
\omega, \sigma)$ from posterior MCMC samples. Stdev is the standard
deviation, and CV is the coefficient of variation.}
\begin{tabular*}
{0.91\textwidth}{cccccccc}
\toprule Time period &  item & Mode &  Mean &  Stdev & Skewness & Kurtosis & CV \\
 \midrule
  & $p$  &  0.0157  & 0.0133 & 0.0022 & 0.951 & 4.86 & 0.168  \\

&  $\rho$ & 0.143 & 0.0623  & 0.0239  & 1.07 & 4.74 & 0.376  \\

  1982-1999 & $\mu$  & 0.471 & 0.456  & 0.027 & 0.221 & 3.64 & 0.058  \\
& $\omega$ & 0.060 & 0.032   & 0.023 &  1.72 & 8.32 & 0.711  \\
& $\sigma$ & 0.448 & 0.457   & 0.085 &  0.912 & 4.50 & 0.183  \\
\midrule
 & $p$  &  0.0177  & 0.0179 & 0.0028 & 0.812 & 4.62 & 0.154  \\

 & $\rho$ & 0.141 & 0.0815  & 0.024  & 1.01 & 4.35 & 0.286  \\

 1982-2010 & $\mu$  & 0.439 & 0.414  & 0.022 & 0.309 & 3.19 & 0.055  \\
& $\omega$ & 0.0717 & 0.031   & 0.016 &  1.24 & 5.39 & 0.51  \\
& $\sigma$ & 0.449 & 0.502   & 0.070 &  0.588 & 3.63 & 0.140  \\
 \bottomrule
\end{tabular*}
\label{tab3}
\end{center}
\end{table}

\begin{table}[!htbp]
\begin{center}
\caption{Summary statistics from posterior MCMC samples for the
stressed PD, LGD and EC, i.e. calculated assuming systematic factor
$x_{T+1}=\Phi^{-1}(0.001)$. $\delta$EC($\%$) is the relative
difference between each quantile value of the distribution of
$Q^{\infty}_\alpha ({{\bm \Theta }})$ calculated using Algorithm 3
and EC value estimated by MLE, $Q^{\infty}_{0.999} ({\hat{\bm \theta
}^{MLE}})$, where $\hat{\bm \theta}^{MLE}$ is the maximum likelihood
estimate for $\bm\theta$.}

\begin{tabular*}
{0.91\textwidth}{cccccccc}
\toprule Time period & item   &  Mean & Stdev &   0.25Q & 0.5Q & 0.75Q &  CV \\
 \midrule
 & PD    & 0.0682 & 0.0236 &  0.0513 & 0.0629 & 0.0800  & 0.346  \\
1982-1999  & LGD    & 0.786 & 0.0745 &   0.733 & 0.777 & 0.829 & 0.0947  \\
 & $Q^{\infty}_{0.999} ({\bm \Theta })$     & 0.0547 & 0.023  & 0.0385 & 0.0489 & 0.0652 & 0.420  \\
& $\delta$EC($\%$)  & 58.1$\%$ &  N/A &  11.3$\%$ & 41.3$\%$ & 88.4$\%$  & N/A  \\
 \midrule
 &  PD    & 0.103 & 0.029 &  0.0825 & 0.0968 & 0.116  & 0.288  \\
1982-2010 &   LGD    & 0.858 & 0.0542 &   0.820 & 0.852 & 0.889 & 0.064  \\
&   $Q^{\infty}_{0.999} ({\bm \Theta })$     & 0.0891 & 0.031  & 0.0683 & 0.0824 & 0.102 & 0.348  \\
& $\delta$EC($\%$)  & 33.8$\%$ &  N/A &  2.55$\%$ & 23.7$\%$ & 53.8$\%$  & N/A  \\
 \bottomrule
\end{tabular*}
\label{tab4}
\end{center}
\end{table}

\begin{table}[!htbp]
\begin{center}
\caption{Full predictive quantile $Q^P_{0.999}$ for various
portfolios (different number of borrowers) using Algorithm 2.}
\begin{tabular*}
{0.65 \textwidth}{ccccc}
\toprule Time period & $J=50$ &  $J=500$ & $J=5000$ &  $J=\infty$  \\
 \midrule
 1982-1999 &  0.1044 & 0.0742 & 0.0732 & 0.0709   \\
 1982-2010 &  0.1454 & 0.1092 & 0.1026 & 0.1026 \\
 \bottomrule
\end{tabular*}
\label{tab5}
\end{center}
\end{table}

\newpage

\end{document}